\documentclass[fleqn,usenatbib]{mnras}

\usepackage{newtxtext,newtxmath}

\usepackage[T1]{fontenc}
\usepackage{ae,aecompl}

\usepackage{graphicx}	
\usepackage{amsmath,mathtools}	
\usepackage{amssymb}	
\usepackage[ruled,vlined]{algorithm2e}  

\DeclareMathAlphabet\mathbfcal{OMS}{cmsy}{b}{n} 

\newcommand{\deriv}{\ensuremath{{\rm d}}}  

\newcommand{\dynesty}{\texttt{dynesty}}
\newcommand{\emcee}{\texttt{emcee}}
\newcommand{\multinest}{\texttt{MultiNest}}
\newcommand{\polychord}{\texttt{PolyChord}}
\newcommand{\nestle}{\texttt{nestle}}

\newcommand{\iid}{\ensuremath{{\rm iid}}}
\newcommand{\Unif}{\ensuremath{{\rm Unif}}}
\newcommand{\Expo}{\ensuremath{{\rm Expo}}}
\newcommand{\Beta}[2]{\ensuremath{{\rm Beta}\left({#1}, {#2}\right)}}

\newcommand{\Normal}[2]{\ensuremath{\mathcal{N}\left({#1}, {#2} \right)}} 

\newcommand{\mean}[1]{\ensuremath{\mathbb{E}\left[{#1}\right]}}
\newcommand{\meanwrt}[2]{\ensuremath{\mathbb{E}_{{#2}}\left[{#1}\right]}}
\newcommand{\variance}[1]{\ensuremath{\mathbb{V}\left[{#1}\right]}}
\newcommand{\stddev}[1]{\ensuremath{\sigma\left[{#1}\right]}}

\newcommand{\params}{\ensuremath{\boldsymbol\Theta}}
\newcommand{\data}{\ensuremath{\mathbf{D}}}
\newcommand{\model}{\ensuremath{M}}
\newcommand{\likelihood}{\ensuremath{\mathcal{L}}}
\newcommand{\prior}{\ensuremath{\pi}}
\newcommand{\posterior}{\ensuremath{\mathcal{P}}}
\newcommand{\evidence}{\ensuremath{\mathcal{Z}}}
\newcommand{\ptform}{\ensuremath{\mathcal{T}}}
\newcommand{\uparams}{\ensuremath{\boldsymbol\Phi}}
\newcommand{\importance}{\ensuremath{\mathcal{I}}}
\newcommand{\stopping}{\ensuremath{\mathcal{S}}}
\newcommand{\cov}{\ensuremath{\mathbf{C}}}

\title[Dynamic Nested Sampling with \dynesty]{\texttt{dynesty}: 
A Dynamic Nested Sampling Package for Estimating Bayesian Posteriors and Evidences}

\author[J. S. Speagle]{
Joshua S. Speagle$^{1,2}$\thanks{E-mail: jspeagle@cfa.harvard.edu}
\\
$^{1}$Center for Astrophysics | Harvard \& Smithsonian, 60 Garden St., Cambridge, MA, USA \\
$^{2}$NSF Graduate Research Fellow
}

\date{Accepted XXX. Received YYY; in original form ZZZ}

\pubyear{2019}

\begin{document}
\label{firstpage}
\pagerange{\pageref{firstpage}--\pageref{lastpage}}
\maketitle

\begin{abstract}
We present \href{https://dynesty.readthedocs.io}{\dynesty}, 
a public, open-source, Python package
to estimate Bayesian posteriors and evidences (marginal likelihoods)
using Dynamic Nested Sampling.
By adaptively allocating samples based on posterior structure,
Dynamic Nested Sampling has the benefits of 
Markov Chain Monte Carlo algorithms that focus exclusively
on posterior estimation while retaining Nested Sampling's
ability to estimate evidences and sample from
complex, multi-modal distributions.
We provide an overview of Nested Sampling,
its extension to Dynamic Nested Sampling,
the algorithmic challenges involved,
and the various approaches taken to solve them.
We then examine {\dynesty}'s performance on a variety of toy problems
along with several astronomical applications. We find in particular problems
{\dynesty} can provide substantial improvements in sampling efficiency
compared to popular MCMC approaches in the astronomical literature.
More detailed statistical results related to Nested Sampling
are also included in the Appendix.
\end{abstract}




\section{Introduction}
\label{sec:intro}

\begin{figure*}
	\includegraphics[width=\textwidth]{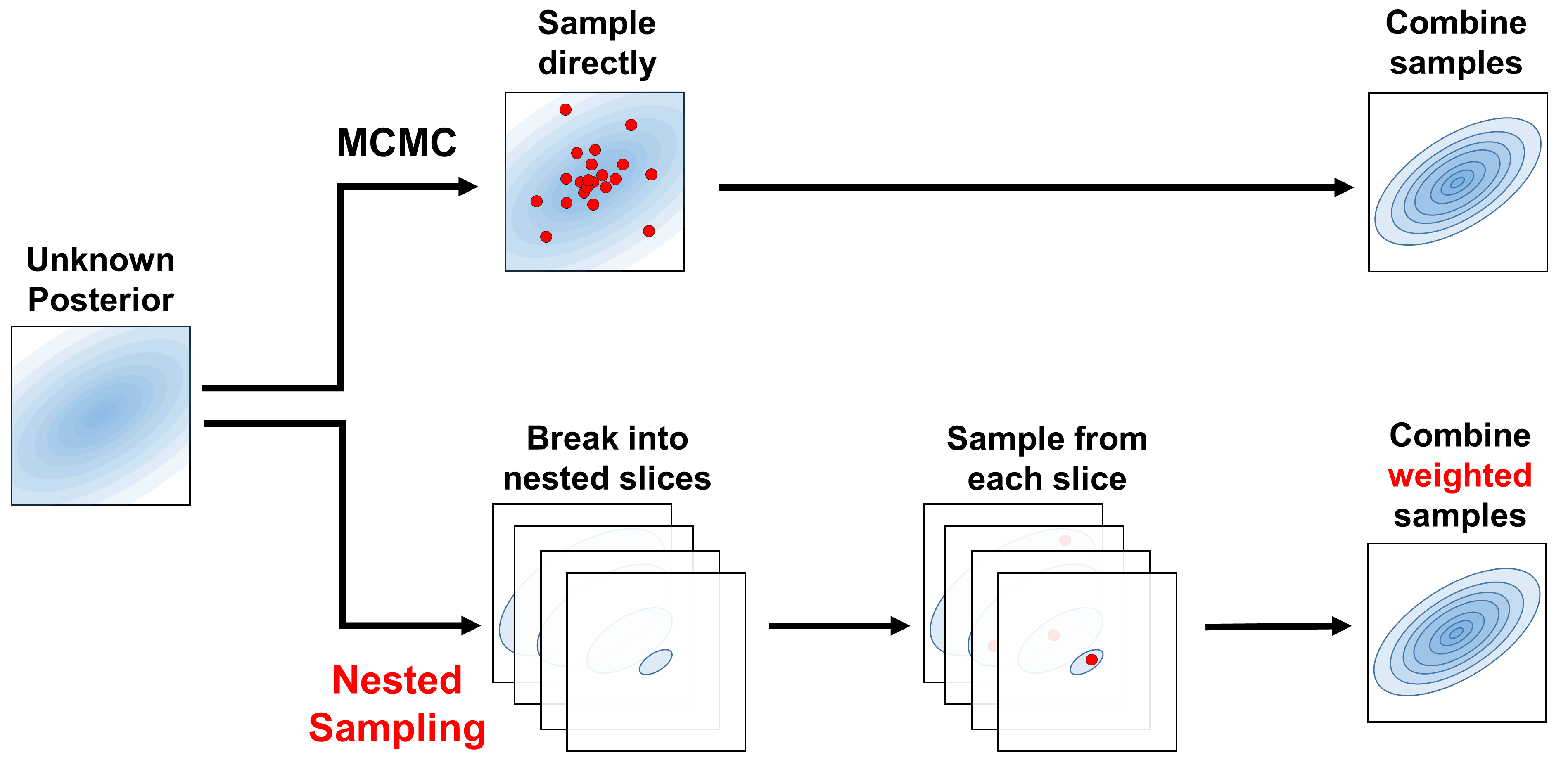}
    \caption{A schematic representation of the different approaches
    Markov Chain Monte Carlo (MCMC) methods and Nested
    Sampling methods take to sample from the posterior. 
    While MCMC methods attempt to generate samples
    directly from the posterior, Nested Sampling instead
    breaks up the posterior into many nested ``slices'',
    generates samples from each of them, and then recombines the
    samples to reconstruct the original distribution using
    the appropriate weights.}
    \label{fig:overview}
\end{figure*}

Much of modern astronomy rests on making inferences about
underlying physical models from observational data.
Since the advent of large-scale, all-sky surveys such 
as SDSS \citep{york+00}, the quality and quantity of these data 
increased substantially \citep{borne+09}. 
In parallel, the amount of computational 
power to process these data also increased enormously. 
These changes opened up an entire new avenue for astronomers
to try and learn about the universe using more complex models
to answer increasingly sophisticated questions over large datasets.
As a result, the standard statistical inference frameworks used in
astronomy have generally shifted away from Frequentist methods
such as maximum-likelihood estimation \citep[MLE;][]{fisher22} to
Bayesian approaches to estimate
the distribution of possible parameters for a given model that are consistent
with the data and our current astrophysical knowledge 
\citep[see, e.g.,][]{trotta08,planck16,fiegelson17}.

In the context of Bayesian inference, we are interested in 
estimating the posterior $P(\params | \data, \model)$ 
of a set of parameters $\params$ for a given model $\model$
conditioned on some data $\data$. 
This can be written into a form commonly known as Bayes Rule
to give
\begin{equation}
    P(\params | \data, \model) 
    = \frac{P(\data | \params, \model) P(\params | \model)}{P(\data | \model)}
\end{equation}
where $P(\data | \params, \model)$ is the likelihood of the
data given the parameters of our model, $P(\params | \model)$ is the prior
for the parameters of our model, and
\begin{equation}
    P(\data | \model)
    = \int_{\Omega_{\params}} P(\data | \params, \model) P(\params | \model) \deriv\params
\end{equation}
is the evidence (i.e. marginal likelihood) for the data given our model,
where the integral is taken over the entire domain $\Omega_{\params}$ of
$\params$ (i.e. over all possible parameter combinations). Throughout the
rest of the paper, we will refer to these using shorthand notation
\begin{equation}
    \posterior(\params_M) = \frac{\likelihood(\params_M) \prior(\params_M)}{\evidence_M}
\end{equation}
where $\posterior(\params_M) \equiv P(\params | \data, \model)$ is the posterior, 
$\likelihood(\params_M) \equiv P(\data | \params, \model)$ is the likelihood,
$\prior(\params_M) \equiv P(\params | \model)$ is the prior,
$\evidence_M \equiv P(\data | \model)$ is the evidence, and
the subscript $M$ will subsequently be dropped if we are only considering a single model.
Here, the posterior $\posterior(\params_M)$ tells us about the parameter estimates
from a \textit{given} model $M$ while $\evidence_M$ enables us to compare \textit{across} models
marginalized over any particular set of parameters using the Bayes factor:
\begin{equation}
    R \equiv \frac{\evidence_{M_1}}{\evidence_{M_2}} \frac{\prior(M_1)}{\prior(M_2)}
\end{equation}
where $\pi(M_i)$ is the prior belief in model $M_i$.

For complicated data and models, the posterior $\posterior(\params)$
is often analytically intractable and must be estimated using numerical
methods. These fall into two broad classes: 
``approximate'' and ``exact'' approaches.
Approximate approaches try to find an (analytic) distribution $\mathcal{Q}(\params)$
that is ``close'' to $\posterior(\params)$ using techniques
such as Variational Inference \citep{blei+16}. 
These techniques are not the focus of this work 
and will not be discussed further in this paper.

Exact approaches try to estimate $\posterior(\params)$ directly,
often by constructing an algorithm that allows us to generate 
a set of samples $\{ \params_1, \params_2, \dots, \params_N \}$ 
that we can use to approximate the posterior as a weighted collection
of discrete points
\begin{equation}
    \posterior(\params) 
    \approx \hat{\posterior}(\params)
    = \frac{\sum_{i=1}^{N} p(\params_i) \delta(\params_i)}
    {\sum_{i=1}^{N} p(\params_i)}
\end{equation}
where $p(\params_i)$ is the importance weight 
associated with each $\params_i$
and $\delta(\params_i)$ is the Dirac delta function located at $\params_i$.

There is a rich literature \citep[see, e.g.,][]{chopinridgway15}
on the approaches used to
generate these samples and their associated weights. The most popular
method used in astronomy today is Markov Chain Monte Carlo (MCMC),
which generates samples ``proportional to'' the posterior
such that $p_i = 1$.
While MCMC has had substantial success over the past few decades 
\citep{brooks+11,sharma17},
the most common implementations 
\citep[e.g.,][]{plummer03,foremanmackey+13,carpenter+17}
tend to struggle when the posterior is comprised of
widely-separated modes. In addition, because it only generates samples
\textit{proportional to} the posterior, it is difficult to use those
samples to estimate the evidence $\evidence_M$ to compare various models.

Nested Sampling \citep{skilling04,skilling06} is an alternative approach to
posterior and evidence estimation
that tries to resolve some of these issues.\footnote{While there are some
hybrid methods that combine Nested Sampling and MCMC 
\citep[e.g., Diffusive Nested Sampling;][]{brewer+09}, we will not discuss
them further here.} By generating samples
in nested (possibly disjoint) ``shells'' of increasing likelihood,
it is able to estimate the evidence $\evidence_M$ for distributions
that are challenging for many MCMC methods to sample from. The final set of
samples can also be combined with their associated importance weights $p_i$
to generate associated estimates of the posterior.\footnote{
While conceptually similar, Nested Sampling is different
from Sequential Monte Carlo (SMC) methods. See \citet{salomone+18} for
additional discussion.}

Since a large portion of modern astronomy relies on being able to perform Bayesian
inference, implementing these methods often can serve as the primary bottleneck
for testing hypotheses, estimating parameters, and performing model comparisons.
As such, packages that implement these approaches serve an important role
enabling science by bridging the gap between writing down a model
and estimating its associated parameters. These allow 
users to perform sophisticated analyses without
having to implement many of the aforementioned algorithms themselves.
Several prominent examples include 
the MCMC package \href{https://emcee.readthedocs.io}{\emcee} \citep{foremanmackey+13}
and the Nested Sampling packages
\href{https://github.com/farhanferoz/MultiNest}{\multinest} \citep{feroz+09,feroz+13}
and \href{https://github.com/PolyChord/PolyChordLite}{\polychord} \citep{handley+15},
which collectively have been used in thousands of papers.

We present \href{https://dynesty.readthedocs.io}{\dynesty}, a
public, open-source, Python package that implements
Dynamic Nested Sampling. {\dynesty} is designed to be 
easy to use and highly modular, with extensive documentation, a straightforward 
application programming interface (API), and a variety of sampling implementations.
It also contains a number of ``quality of life'' features including well-motivated
stopping criteria, plotting functions, and analysis utilities
for post-processing results.

The outline of the paper is as follows.
In \S\ref{sec:nested} we give an overview of
Nested Sampling and discuss the method's benefits and drawbacks.
In \S\ref{sec:dynamic} we describe how Dynamic Nested Sampling
is able to resolve some of these drawbacks by allocating samples more flexibly.
In \S\ref{sec:methods} we discuss the specific approaches
{\dynesty} uses to track and sample from complex, multi-modal
distributions.
In \S\ref{sec:tests} we examine \dynesty's performance on a variety
of toy problems.
In \S\ref{sec:applied} we examine \dynesty's performance on several
real-world astrophysical analyses.
We conclude in \S\ref{sec:conclusion}.
For interested readers, more detailed results on many of the
methods outlined in the main text are included in Appendix \ref{ap:nested}.

{\dynesty} is publicly available on 
\href{https://github.com/joshspeagle/dynesty}{GitHub}
as well as on \href{https://pypi.org/project/dynesty/}{PyPI}.
See \url{https://dynesty.readthedocs.io} for installation 
instructions and examples on getting started.

\section{Nested Sampling}
\label{sec:nested}

The general motivation for Nested Sampling, first
proposed by \citet{skilling04} and later 
fleshed out in \citet{skilling06}, stems from the fact that
sampling from the posterior $\posterior(\params)$ directly is \textit{hard}.
Methods such as Markov Chain Monte Carlo (MCMC) attempt to tackle this
single difficult problem \textit{directly}. Nested Sampling, however,
instead tries to break down this single hard problem
into a larger number of \textit{simpler} problems by:
\begin{enumerate}
    \item ``slicing'' the posterior into many
    simpler distributions,
    \item sampling from each of those in turn, and
    \item re-combining the results afterwards.
\end{enumerate}
We provide a schematic
illustration of this procedure in Figure \ref{fig:overview} and
give a broad overview of this process below.
For additional details, please see Appendix \ref{ap:nested}.

\begin{algorithm*}
    \SetAlgoLined
    \tcp{Initialize live points.}
    Draw $K$ ``live'' points
    $\lbrace \params_1, \dots, \params_K \rbrace$ 
    from the prior $\prior(\params)$. \\
    \tcp{Main sampling loop.}
    \While{${\rm stopping\:criterion\:not\:met}$}{
        Compute the minimum likelihood $\likelihood^{\min}$ among the
        current set of live points. \\
        Add the $k$th live point $\params_k$ associated with $\likelihood^{\min}$
        to a list of ``dead'' points. \\
        Sample a new point $\params'$ from the prior subject to the constraint
        $\likelihood(\params') \geq \likelihood^{\min}$. \\
        Replace $\params_k$ with $\params'$. \\
        \tcp{Check whether to stop.}
        Evaluate stopping criterion. \\
    }
    \tcp{Add final live points.}
    \While{$K > 0$}{
        Compute the minimum likelihood $\likelihood^{\min}$ among the
        current set of live points. \\
        Add the $k$th live point $\params_k$ associated with $\likelihood^{\min}$
        to a list of ``dead'' points. \\
        Remove $\params_k$ from the set of live points. \\
        Set $K = K - 1$. \\
    }
    \caption{Static Nested Sampling}
    \label{alg:static}
\end{algorithm*}

\subsection{Overview}
\label{subsec:nested_overview}

Unlike MCMC methods, which attempt to estimate the posterior $\posterior(\params)$
directly, Nested Sampling instead focuses on estimating the evidence
\begin{equation}
    \evidence 
    \equiv \int_{\Omega_{\params}} \posterior(\params) \deriv\params
    = \int_{\Omega_{\params}} \likelihood(\params) \prior(\params) \deriv\params
\end{equation}
As this integral is over the entire multi-dimensional domain of $\params$,
it is traditionally very challenging to estimate. 

Nested Sampling approaches this problem by re-factoring this integral
as one taken over prior volume $X$ of the enclosed parameter space
\begin{equation} \label{eq:ns_evid}
    \evidence = \int_{\Omega_{\params}} \likelihood(\params) \prior(\params) \deriv\params
    = \int_{0}^{1} \likelihood(X) \deriv X
\end{equation}
Here, $\likelihood(X)$ now defines an iso-likelihood contour (or multiple)
defining the edge(s) of the volume $X$, while the prior volume
\begin{equation}
    X(\lambda) 
    \equiv \int_{\Omega_{\params}: \likelihood(\params) \geq \lambda} \prior(\params) \deriv\params
\end{equation}
is the fraction of the prior where the likelihood $\likelihood(\params) \geq \lambda$ is above
some threshold $\lambda$. Since the prior is normalized, this gives $X(\lambda=0)=1$ and
$X(\lambda=\infty)=0$, which define the bounds of integration for equation \eqref{eq:ns_evid}.

As a rough analogy, we can consider trying to integrate over a spherically-symmetric
distribution in 3-D. While it is possible to integrate over $\deriv x \deriv y \deriv z$
directly, it often is significantly easier to instead integrate
over differential volume elements $\deriv V=4 \pi r^2$ as a function of radius 
$r \equiv \sqrt{x^2 + y^2 + z^2}$:
\begin{equation*}
    \int \posterior(x,y,z) \deriv x \deriv y \deriv z 
    = \int \posterior(V(r)) \deriv V(r)
    = \int \posterior(r) 4\pi r^2 \deriv r
\end{equation*}
Parameterizing the evidence integral
this way allows Nested Sampling (in theory)
to convert from a complicated $D$-dimensional integral over $\params$
to a simple 1-D integral over $X$.

While it is straightforward to
evaluate the likelihood at a given position $\likelihood(\params)$,
estimating the associated prior volume $X(\params)$ 
and its differential $dX(\params)$ is substantially more challenging.
We can, however, generate \textit{noisy estimates} of these quantities
by employing the procedure described in Algorithm \ref{alg:static}.
We elaborate further on this procedure and how it works below.

\subsection{Generating Samples}
\label{subsec:nested_samples}

A core element of Nested Sampling is the ability to generate samples
from the prior $\prior(\params)$ subject to a
hard likelihood constraint $\lambda$.
The most naive algorithm that satisfies this constraint is
simple rejection sampling: at a given iteration $i$,
generate samples $\params_{i+1}$ from the prior $\prior(\params)$ until
$\likelihood(\params_{i+1}) \geq \likelihood(\params_i)$.

In practice, however, this simple procedure 
becomes progressively less efficient as time goes on
since the remaining prior volume $X_{i+1}$ at each iteration
of Algorithm \ref{alg:static} keeps shrinking. We therefore
need a way of directly generating samples from the
constrained prior:
\begin{equation}
    \prior_\lambda(\params) 
    \equiv
    \begin{cases}
    \prior(\params) / X(\lambda) & \likelihood(\params) \geq \lambda \\
    0 & \likelihood(\params) < \lambda
    \end{cases}
\end{equation}

Sampling from this constrained distribution
is difficult for an arbitrary prior $\prior(\params)$
since the density can vary drastically from place to place.
It is simpler, however, if the prior is
standard uniform (i.e. flat from $0$ to $1$)
in all dimensions so that the density interior to
$\lambda$ is constant then $X$ behaves more like 
a typical volume $V$. We can accomplish this through the
use of the appropriate ``prior transform'' function $\ptform$
which maps a set of parameters $\uparams$
with a uniform prior over the $D$-dimensional
unit cube to the parameters of interest $\params$.\footnote{
In general, there is a uniquely defined prior transform $\ptform$ for
any given $\prior(\params)$; see the {\dynesty}
\href{https://dynesty.readthedocs.io/en/latest/quickstart.html}{documentation}
for additional details.}
Taken together, these transform our 
original hard problem of sampling from the posterior
$\posterior(\params)$ directly to instead 
the much simpler problem of repeatedly sampling 
uniformly\footnote{Technically this requirement 
is overly strict, as Nested Sampling
can still be valid even if the samples at each iteration are correlated.
See Appendix \ref{ap:nested} for additional discussion.} within the
transformed constrained prior
\begin{equation}
    \prior'_\lambda(\uparams) 
    \equiv
    \begin{cases}
    1 / X(\lambda) & \likelihood(\params = \ptform(\uparams)) \geq \lambda \\
    0 & {\rm otherwise}
    \end{cases}
\end{equation}
Throughout the rest of the text we will henceforth assume $\pi(\params)$
is a unit cube prior unless otherwise explicitly specified.

Because there is no constraint that
this distribution is uni-modal, the constrained prior
may define several ``blobs'' of prior volume that we
are interested in sampling from.
While sampling from the blob(s) 
might be hard to do from scratch, because Nested Sampling 
samples at many different likelihood ``levels'',
structure tends to emerge over time rather than all at once
as we transition away from the prior $\prior(\params)$.

\begin{figure*}
	\includegraphics[width=\textwidth]{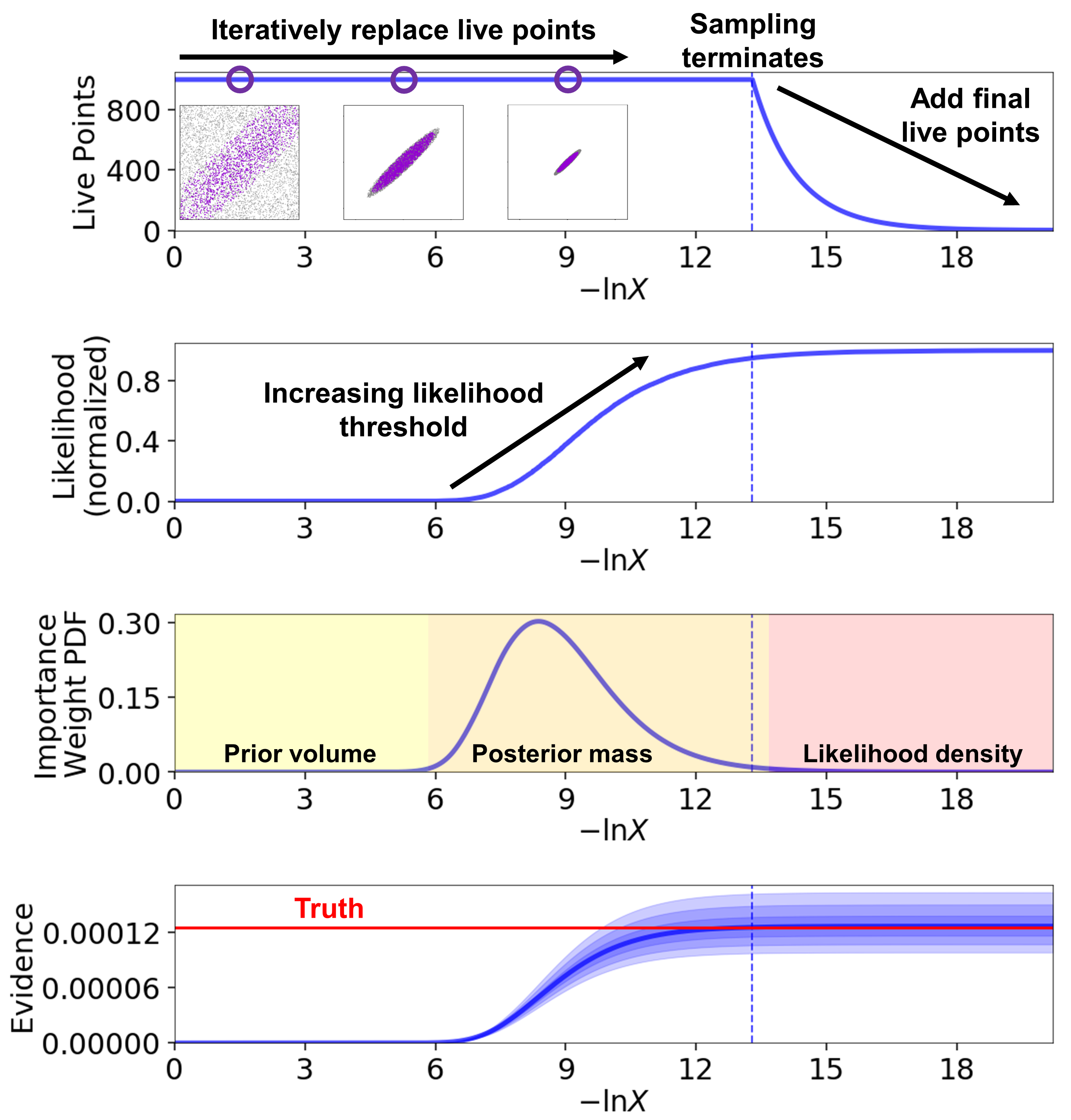}
    \caption{An example highlighting the behavior
    of a Static Nested Sampling run in {\dynesty}.
    See \S\ref{sec:nested} for additional details. 
    \textit{Top:} The number of live points as
    a function of prior volume $X$. Snapshots of their
    their distribution (purple) with respect
    to the current bounds (gray; see \S\ref{subsec:bounds}) 
    are highlighted in several insets. The number of live
    points remains constant until sampling terminates,
    at which point we add the final live points one-by-one
    to the samples. 
    \textit{Top-middle:} The (normalized) likelihood limit 
    $\likelihood/\likelihood^{\max}$ associated with a 
    the prior volume $X(\likelihood)$ in the top panel.
    This increases monotonically as we sample increasingly
    smaller regions of the prior. 
    \textit{Bottom-middle:} The importance
    weight PDF $p(X)$, roughly 
    divided into regions dominated by the prior volume ($\deriv X$ is 
    large, $\likelihood(X)$ is small; yellow),
    posterior mass ($\deriv X$ and $\likelihood(X)$ are 
    comparable; orange), and likelihood density 
    ($\deriv X$ is small, $\likelihood(X)$
    is large; red). The posterior mass is the 
    most important for posterior estimation, while
    evidence estimation also depends on the prior volume. 
    \textit{Bottom:} The estimated evidence $\hat{\evidence}(X)$
    (blue line) and its 1, 2, and 3-sigma errors (blue shaded).
    The true value is shown in red.}
    \label{fig:static}
\end{figure*}

\subsection{Estimating the Prior Volume}
\label{subsec:nested_volume}

As shown in Appendix \ref{ap:nested}, generating samples following the
strategy in \S\ref{subsec:nested_samples} 
based on Algorithm \ref{alg:static} allows us to estimate
the (change in) prior volume at a given iteration using the set of ``dead'' points
(i.e. the live points we replaced at each iteration). In particular, it
leads to \textit{exponential shrinkage} such that the (log-)prior volume 
at each iteration changes by
\begin{equation}
    \mean{\Delta \ln \hat{X}_i} 
    = \mean{\ln \hat{X}_{i} - \ln \hat{X}_{i-1}} 
    = -\frac{1}{K}
\end{equation}
where $\mean{\cdot}$ is the expectation value (i.e. mean) and we have adopted
the $\hat{x}$ notation to emphasize that we have a noisy estimator of
the prior volume $X$. Using more live points $K$ thus increases our volume
resolution by decreasing the rate of this exponential compression.
By default, {\dynesty} uses $K=500$ live points, although this should
be adjusted depending on the problem at hand.

Once some stopping criterion is reached and sampling terminates
after $N$ iterations, 
the remaining set of $K$ live points are then distributed 
uniformly within the final prior volume $X_N$ (see Appendix \ref{ap:nested}).
These can be ``recycled'' into the final set of samples by sequentially
adding the live points to the list of ``dead'' points collected at each iteration
in order of increasing likelihood. This leads to
\textit{uniform shrinkage} of the prior volume such that the
(fractional) change in prior volume for the $k$th live point added this way is
\begin{equation}
    \mean{\frac{\Delta \hat{X}_{N+k}}{\hat{X}_N}}
    = \mean{\frac{\hat{X}_{N+k} - \hat{X}_{N+k-1}}{\hat{X}_N}} 
    = \frac{1}{K+1}
\end{equation}
where $\hat{X}_N$ is the estimating remaining prior volume at the final
$N$th iteration.

\subsection{Stopping Criterion}
\label{subsec:nested_stop}

Since Nested Sampling is designed to estimate the evidence, a natural
stopping criterion \citep[see, e.g.,][]{skilling06,keeton11,higson+17a}
is to terminate sampling when we believe
our set of dead points (and optionally the remaining live points) give us an
integral that encompasses the vast majority of the posterior.
In other words, at a given iteration $i$, we want to terminate sampling if
\begin{equation}
    \Delta\ln\hat{\evidence}_i 
    \equiv \ln\left(\hat{\evidence}_i + \Delta\hat{\evidence}_i\right)
    - \ln\left(\hat{\evidence}_i\right) < \epsilon
\end{equation}
where $\Delta\hat{\evidence}_i$ is the estimated remaining evidence
we have yet to integrate over and $\epsilon$ determines the tolerance.
If the final set of live points are excluded from the set of dead points,
{\dynesty} assumes a default value of $\epsilon = 10^{-2}$
(i.e. $\lesssim1\%$ of the evidence remaining). If the final set of
live points are included, {\dynesty} instead uses the slightly
more permissive $\epsilon = 10^{-3}(K-1) + 10^{-2}$.

While the remaining evidence $\Delta\hat{\evidence}_i$ is unknown,
we can in theory construct a strict upper bound on it by assigning
\begin{equation}
    \Delta\hat{\evidence}_i \leq \likelihood^{\max} X_i
\end{equation}
where $\likelihood^{\max}$ is the maximum-likelihood value
across the entire domain $\Omega_{\params}$ and $X_i$ is the
prior volume at the current iteration. This is equivalent to treating the remaining
likelihood interior to the current sample ($X < X_i$) as a uniform slab
with amplitude $\likelihood^{\max}$.

Unfortunately, neither $\likelihood^{\max}$ or $X_i$ is known exactly. However,
we can approximate this upper bound by replacing both quantities with
associated estimators to get the \textit{rough} upper bound
\begin{equation}
    \Delta\hat{\evidence}_i \lesssim \likelihood_i^{\max} \hat{X}_i
\end{equation}
where $\likelihood_i^{\max}$ is the maximum value of the likelihood
among the live points at iteration $i$ and $\hat{X}_i$ is the estimated
(remaining) prior volume.

While this rough upper bound works well in most cases, because 
we only have access to the best likelihood $\likelihood_i^{\max}$
sampled by the $K$ live points at a particular iteration
there is always a chance that $\likelihood_i^{\max} \ll \likelihood^{\max}$
and that we will terminate early. This can happen if there is an extremely
narrow likelihood peak within the remaining prior volume that has
not yet been discovered by the $K$ live points.

\subsection{Estimating the Evidence and Posterior}
\label{subsec:nested_outputs}

Once we have a final set of samples $\lbrace \params_1, \dots, \params_N \rbrace$, 
we can estimate the 1-D evidence integral using standard numerical techniques.
To ensure approximation errors on the numerical integration
estimate are sufficiently small, {\dynesty} uses the 2nd-order trapezoid rule
\begin{equation}
    \hat{\evidence} 
    = \sum_{i=1}^{N+K} 
    \frac{1}{2} \left[\likelihood(\params_{i-1}) + \likelihood(\params_i)\right]
    \times \left[\hat{X}_{i-1} - \hat{X}_i\right] \equiv \sum_{i=1}^{N+K} \hat{p}_i
\end{equation}
where $X_0=X(\lambda=0)=1$ and 
\begin{equation}
    \hat{p}_i 
    \equiv \left[\likelihood(\params_{i-1}) + \likelihood(\params_i)\right]
    \times \left[\hat{X}_{i-1} - \hat{X}_i\right]
\end{equation}
is the estimated importance weight.
By default, {\dynesty} uses the mean values of $\hat{X}_i$ to
compute the mean and standard deviation of $\ln\hat{\evidence}$
following Appendix \ref{ap:nested}, although these values can also be simulated
explicitly.

We can also estimate the posterior $\posterior(\params)$
from the same set of $N+K$ dead points by using the associated
importance weights derived above:
\begin{equation}
    \hat{\posterior}(\params)
    = \frac{\sum_{i=1}^{N+K} \hat{p}(\params_i) \delta(\params_i)}
    {\sum_{i=1}^{N+K} \hat{p}(\params_i)}
    = \hat{\evidence}^{-1} \sum_{i=1}^{N+K} \hat{p}(\params_i) \delta(\params_i)
\end{equation}
By default, {\dynesty} uses the mean values of $\hat{X}_i$ to compute this
posterior estimate, although as with the evidence these values can also be simulated
explicitly (see Appendix \ref{ap:nested}).

An illustration of a typical Nested Sampling run is shown in Figure \ref{fig:static}.

\subsection{Benefits of Nested Sampling}
\label{subsec:nested_good}

Because of its alternative approach to sampling from the posterior, Nested Sampling
has a number of benefits relative to traditional MCMC approaches:
\begin{enumerate}
    \item Nested Sampling can estimate the evidence $\evidence$
    as well as the posterior $\posterior(\params)$. MCMC methods generally can only
    constrain the latter \citep[although see][]{lartillotphilippe06,heavens+17}.
    \item Nested sampling can sample from multi-modal distributions
    that tend to challenge many MCMC methods.
    \item While most MCMC stopping criteria based on effective sample sizes
    can feel arbitrary, Nested Sampling possesses
    well-motivated stopping criteria focused on 
    evidence estimation.
    \item MCMC methods need to converge (i.e. ``burn in'')
    to the posterior before any samples
    generated are valid. While optimization techniques can
    speed up this process, assessing this convergence
    can be challenging and time-consuming \citep{gelmanrubin92,vehtari+19}. 
    Nested Sampling doesn't suffer from similar issues
    because the method smoothly integrates 
    over the posterior $\posterior(\params)$
    starting from the prior $\prior(\params)$.
\end{enumerate}

\subsection{Drawbacks}
\label{subsec:nested_bad}

While Nested Sampling has its fair share of benefits that have encouraged its
rapid adoption in astronomical Bayesian analyses,
it also suffers from a fair share of drawbacks.
Most crucially, the standard Nested Sampling implementation outlined in
Algorithm \ref{alg:static} focuses \textit{exclusively} on estimating
the evidence $\evidence$; the posterior $\posterior(\params)$ is entirely
a by-product of the approach. This creates several immediate drawbacks relative to
MCMC, which focuses exclusively on sampling the posterior $\posterior(\params)$.


First, because most Nested Sampling implementations rely on sampling
from uniform distributions
(see \S\ref{subsec:nested_samples}), applying them to general distributions
requires knowing the appropriate prior transform $\ptform$.
While these are straightforward
to define when the prior can be decomposed into separable, independent
components, they can be more difficult to derive when the prior involves
conditional and/or jointly distributed parameters.

\begin{algorithm*}
    \SetAlgoLined
    \tcp{Initialize first set of live points.}
    Draw $K$ ``live'' points
    $\lbrace \params_1, \dots, \params_K \rbrace$ 
    from the prior $\prior(\params)$. \\
    \tcp{Main sampling loop.}
    Set $\likelihood^{\min} = 0$ and $K_0 = K$. \\
    \While{${\rm stopping\:criterion\:not\:met}$}{
        \tcp{Get current number of live points.}
        Compute the previous number of live points $K$ and the
        current number of live points $K'$. \\
        \eIf{$K' \geq K$}{
            \tcp{Add in new live points.}
            \While{$K' > K$}{
                Sample a new point $\params'$ from the prior 
                subject to the constraint
                $\likelihood(\params') \geq \likelihood^{\min}$. \\
                Add $\params'$ to the set of live points. \\
                Set $K = K + 1$. \\
            }
            \tcp{Replace worst live point.}
            Compute the minimum likelihood $\likelihood^{\min}$ among the
            current set of $K$ live points. \\
            Add the $k$th live point $\params_k$
            associated with $\likelihood^{\min}$
            to a list of ``dead'' points. \\
            Replace $\params_k$ with $\params'$. \\
        }{
            \tcp{Iteratively remove live points.}
            \While{$K' < K$}{
                Compute the minimum likelihood $\likelihood^{\min}$ among the
                current set of $K=K'$ live points. \\
                Add the $k$th live point $\params_k$ 
                associated with $\likelihood^{\min}$
                to a list of ``dead'' points. \\
                Remove $\params_k$ from the set of live points. \\
                Set $K = K - 1$.
            }
        }
        \tcp{Check whether to stop.}
        Evaluate stopping criterion. \\
    }
    \tcp{Add final live points.}
    \While{${\rm there\:are\:live\:points\:remaining}$}{
        Compute the minimum likelihood $\likelihood^{\min}$ among the
        current set of live points. \\
        Add the $k$th live point $\params_k$ 
        associated with $\likelihood^{\min}$
        to a list of ``dead'' points. \\
        Remove $\params_k$ from the set of live points. \\
    }
    \caption{Dynamic Nested Sampling}
    \label{alg:dynamic}
\end{algorithm*}

Second, because the evidence depends on the 
amount of prior volume that needs to be integrated over,
the overall expected runtime is sensitive 
to the relative size of the prior.
In other words, while estimating the posterior 
mostly depends on generating samples
close to where the majority of the distribution 
is located \citep[i.e. the ``typical set'';][]{betancourt+17},
estimating the evidence requires generating 
samples in the extended tails of the distribution.
Using less informative (broader) priors
will increase the expected runtime even 
if the posterior is largely unchanged.

Finally, because the number of live points 
$K$ is constant, the rate $\Delta \ln X$
at which we integrate over the posterior $\posterior(\params)$ 
is the same regardless
of where we are. This means that increasing the 
number of like points $K$, which
increases the overall runtime, always improves the 
accuracy of \textit{both} the posterior $\hat{\posterior}(\params)$
and evidence $\hat{\evidence}$ estimates.
In other words, Nested Sampling does 
not allow users to \textit{prioritize} between
estimating the posterior or the evidence,
which is not ideal for many analyses 
that are mostly interested in using Nested 
Sampling for either option. We focus on improving this
behavior in \S\ref{sec:dynamic}.

As with any sampling method, we strongly advocate that 
Nested Sampling \textit{should not} be viewed as being
strictly ``better'' or ``worse'' than MCMC, but 
rather as a tool that can be more 
or less useful in certain problems.
There is no ``One True Method to Rule Them All'',
even though it can be tempting to look for one.

\section{Dynamic Nested Sampling}
\label{sec:dynamic}

In our overview of Nested Sampling in \S\ref{sec:nested}, we 
highlighted three main drawbacks of basic implementations:
\begin{enumerate}
    \item They generally require a prior transform.
    \item Their runtime is sensitive to the size of the prior.
    \item Their rate of posterior integration is always constant.
\end{enumerate}
While the first two drawbacks are essentially
inherent to Nested Sampling as sampling strategy, the last is not.
Instead, the inability of Algorithm \ref{alg:static} to ``prioritize''
estimating the evidence $\evidence$ or posterior $\posterior(\params)$
is a consequence of the fact that the number of 
live points $K$ remains constant
throughout an entire run, which sets 
the rate of integration $\Delta \ln X$.
As a result, we will henceforth call this
procedure ``Static'' Nested Sampling.

To address this issue, \citet{higson+17b}
proposed a deceptively simple modification:
let the number of live points \textit{vary} during runtime.
This gives a new ``Dynamic'' Nested Sampling
algorithm whose basic implementation is
outlined in Algorithm \ref{alg:dynamic}. This
simple change is transformative, allowing Dynamic Nested Sampling
to focus on sampling the posterior
$\posterior(\params)$, similar to MCMC approaches,
while retaining all the benefits of (Static) Nested Sampling to
estimate the evidence $\evidence$ and sample from complex, multi-modal
distributions. It also possesses well-motivated new stopping criteria
for posterior and evidence estimation.

It is important to note that we cannot take advantage 
of the flexibility offered by Dynamic Nested Sampling, however,
without implementing appropriate schemes
to specify exactly how live points should be allocated, when to
terminate sampling, etc. While {\dynesty}
tries to implement a number of reasonable default choices, in practice
this inevitably leads to many more tuning parameters that can affect the behavior
of a given Dynamic Nested Sampling run.

We provide an illustration of the overall 
approach in Figure \ref{fig:dynamic}
and give a broad overview of the basic algorithm below.
For additional details, please see Appendix \ref{ap:nested}.

\begin{figure*}
	\includegraphics[width=\textwidth]{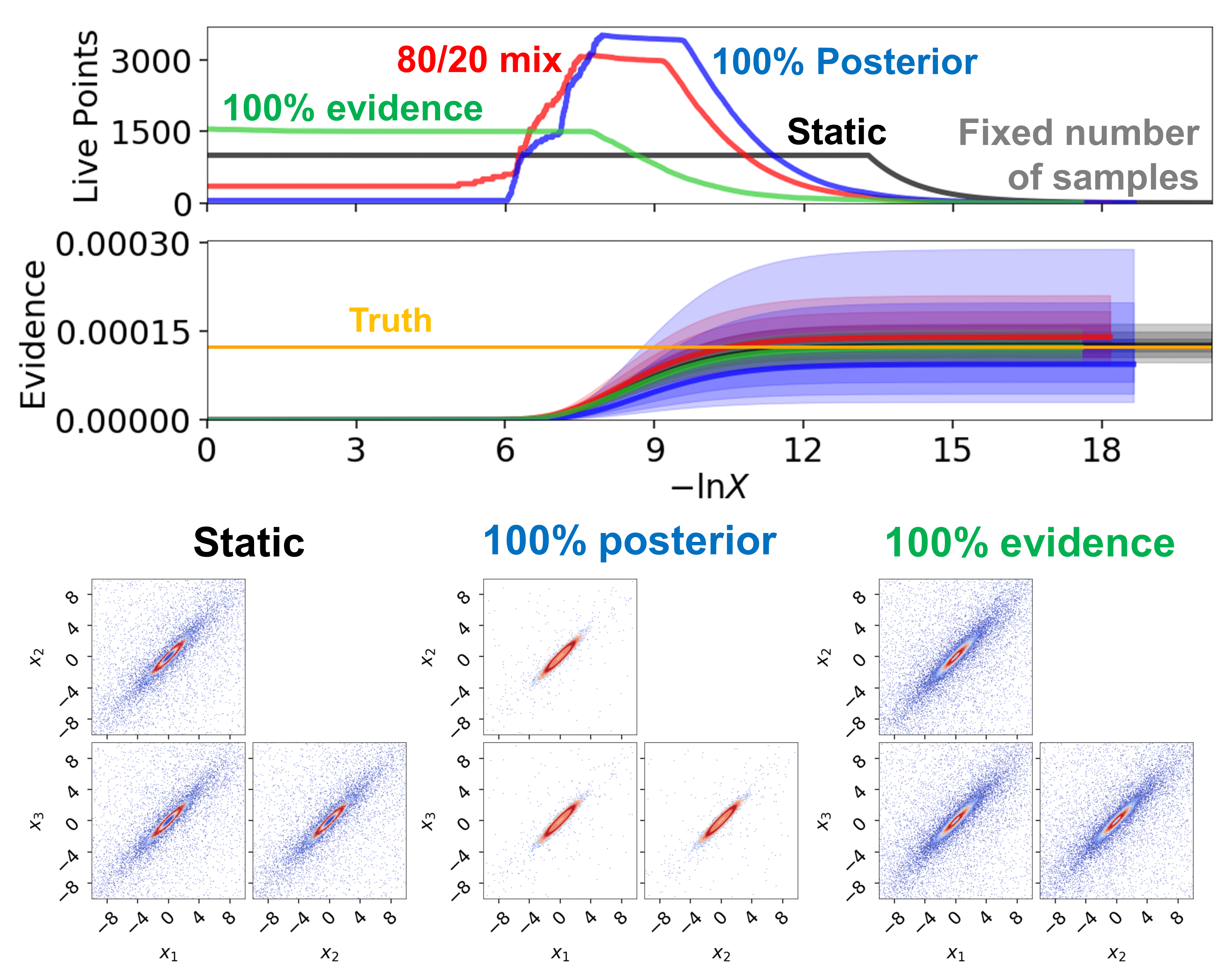}
    \caption{An example highlighting different
    schemes for live point allocation between Static 
    and Dynamic Nested Sampling run in {\dynesty}
    \textit{with a fixed number of samples}.
    See \S\ref{sec:dynamic} for additional details. 
    \textit{Top panels:} As Figure \ref{fig:static},
    but now highlighting the number of live points
    (upper) and evidence estimates (lower) for
    a Static Nested Sampling run (black) and Dynamic
    Nested Sampling runs focused entirely on estimating
    the posterior (blue), entirely on estimating the evidence
    (green), and with an 80\%/20\% posterior/evidence
    mixture (the default in {\dynesty}; red). 
    \textit{Bottom panels:} The distribution of samples
    from the targeted 3-D correlated Gaussian distribution
    in the Static (left), posterior-focused (middle),
    and evidence-focused (right) runs. Points are
    color-coded based on their important weight $p_i$.
    The posterior-oriented run allocates points almost exclusively 
    around the bulk of the posterior mass, while the
    evidence-oriented run preferentially allocates them in
    prior-dominated regions.
    }
    \label{fig:dynamic}
\end{figure*}

\subsection{Allocating Live Points}
\label{subsec:dynamic_live}

The singular defining feature of the Dynamic Nested Sampling algorithm
is the scheme we use for determining how the number
of live points $K_i$ at a given iteration $i$ should vary.
Naively, we would like $K_i$ to be larger where we want 
our resolution to be higher (i.e. a slower rate of integration
$\Delta \ln X_i$) and smaller where we are interested in
traversing the current region of prior volume more quickly. This
allows us to prioritize adding samples in regions of interest.

In general, we would like the number of live points 
$K(X)$ as a function of prior volume $X$ to follow
a particular importance function $\importance(X)$ such that
\begin{equation}
    K(X) \propto \importance(X)
\end{equation}
While this function can be completely general, since most users
are interested in estimating the posterior
$\posterior(\params)$ and/or evidence $\evidence$ more
generally, {\dynesty} by default follows \citet{higson+17b} and
considers a function of the form:
\begin{equation}
    \importance(X) 
    = f^{\posterior} \importance^{\posterior}(X)
    + (1 - f^{\posterior}) \importance^{\evidence}(X)
\end{equation}
where $f^{\posterior}$ is the relative amount of importance
placed on estimating the posterior.

We define the posterior importance function as
\begin{equation}
    \importance^{\posterior}(X) \equiv p(X)
\end{equation}
where $p(X)$ is the now the probability density function (PDF)
of the importance weight defined in \S\ref{subsec:nested_outputs}.
This choice just means that we want to allocate more live points
in regions where the posterior mass $\propto \likelihood(X) \deriv X$ is
higher.

We define the evidence importance function as
\begin{equation}
    \importance^{\evidence}(X) 
    \equiv \frac{1 - \evidence(X)/\evidence}
    {\int_0^1 (1 - \evidence(X)/\evidence) \deriv X}
\end{equation}
where $\evidence(X)$ is the evidence integrated \textit{up to} $X$.
This means that we want to allocate more live points
when we believe we have not integrated over 
much of the posterior (i.e. in the prior volume-dominated
regime at larger values of $X$) and fewer as we
integrate over larger portions of the posterior mass
and become more confident in our estimated value of $\evidence$
(see Figure \ref{fig:static}).

\begin{algorithm*}
    \SetAlgoLined
    \tcp{Baseline Nested Sampling run.}
    Run Static Nested Sampling (Algorithm \ref{alg:static}) with: \\
    (a) $K$ live points \\
    (b) sampled uniformly from the prior $\prior(\params)$ \\
    (c) until the default Static Nested Sampling stopping criterion is met. \\
    \tcp{Main sampling loop.}
    \While{${\rm stopping\:criterion\:not\:met}$}{
        \tcp{Find region where new samples should be allocated.}
        Compute relative importance 
        $\lbrace \hat{\importance}(\hat{X}_i) \rbrace$ 
        over all dead points $\lbrace \params_i \rbrace$. \\
        Use $\lbrace \hat{\importance}_i \rbrace$ to assign
        lower $\likelihood^{\rm low} = \likelihood(\hat{X}^{\rm high})$ and 
        upper $\likelihood^{\rm high} = \likelihood(\hat{X}^{\rm low})$ 
        likelihood bounds. \\
        \tcp{Batch Nested Sampling run.}
        Run Static Nested Sampling (Algorithm \ref{alg:static}) with: \\
        (a) $K'$ live points \\
        (b) sampled uniformly from the constrained prior $\prior_\lambda(\params)$
        based on the lower likelihood bound $\lambda=\likelihood^{\rm low}$ \\
        (c) until the likelihood $\likelihood(\params)$ 
        of the last dead point
        exceeds the upper likelihood bound $\likelihood^{\rm high}$. \\
        \tcp{Merge samples from batch.}
        Merge new batch of dead points $\lbrace \params_i' \rbrace$
        into the previous set of dead points 
        $\lbrace \params_i \rbrace$. \\
        \tcp{Check whether to stop.}
        Evaluate stopping criterion. \\
    }
    \caption{Iterative Dynamic Nested Sampling}
    \label{alg:dynamic_iter}
\end{algorithm*}

\subsection{Iterative Dynamic Nested Sampling}
\label{subsec:dynamic_iter}

As in \S\ref{subsec:nested_stop}, we unfortunately do not have
access to $X$ or $\importance(X)$ directly. 
We thus need to use noisy estimators
to approximate them, which are only available 
\textit{after we have already generated samples
from the posterior}. In practice then, Dynamic Nested Sampling
works as an iterative modification to Static Nested Sampling.
We outline this ``Iterative'' Dynamic Nested Sampling 
approach, first proposed in \citet{higson+17b}
and implemented in {\dynesty}, in Algorithm \ref{alg:dynamic_iter}.
It has five main steps:
\begin{enumerate}
    \item Sample the distribution with Static Nested Sampling 
    to lay down a ``baseline run'' to get a sense where
    the posterior mass $\posterior(X) \deriv X$ is located.
    \item Evaluate our importance function $\importance(X)$
    over the existing set of samples.
    \item Use the computed importances $\importance_i$ 
    to decide where to allocate additional live points/samples.
    \item Add a new ``batch'' of samples in the 
    region of interest using Static Nested Sampling.
    \item ``Merge'' the new batch of samples 
    into the previous set of samples.
\end{enumerate}
We then repeat steps (ii) to (v) until some stopping criterion is met.
By default, {\dynesty} uses $K_{\rm base}=K_{\rm batch}=250$
points for each run, although this should be
adjusted depending on the problem at hand.

Allocating points using an existing set of
samples is a two-step process. First, we evaluate
a noisy estimate of our importance function over the samples:
\begin{equation}
    \hat{\importance}_i 
    = f^{\posterior} \frac{\hat{p}_i}{\sum_{i=1}^{N} \hat{p}_i} 
    + (1 - f^{\posterior}) 
    \frac{1 - \hat{\evidence}_i/(\hat{\evidence}_N + \Delta \hat{\evidence}_N)}
    {\sum_{i=1}^{N} 1 - \hat{\evidence}_i/(\hat{\evidence}_N + \Delta \hat{\evidence}_N)}
\end{equation}
where we are now using the noisy importance weight 
$\hat{p}_i$ to estimate the posterior
and the rough upper limit
$\Delta \hat{Z}_N \sim \likelihood^{\max}_N \hat{X}_N$ to estimate the
remaining evidence. Then, we use these values to 
define new regions of prior
volume to sample. By default, {\dynesty} only samples 
from a single contiguous range
of prior volume $(X^{\rm low}, X^{\rm high}]$ 
which define an associated
(flipped) range in iteration $[i^{\rm low}, i^{\rm high})$ and 
likelihood $[\likelihood^{\rm low}, \likelihood^{\rm high})$ 
defined by the simple heuristic
\begin{align}
    i^{\rm low} 
    &= \min\left[\min(\lbrace i \rbrace) - n_{\rm pad}, 0\right] \nonumber \\
    i^{\rm high} 
    &= \max\left[\max(\lbrace i \rbrace) + n_{\rm pad}, N\right] \\
    &\forall \: i \in [0, N] \:\:{\rm s.t.}\:\: \hat{\importance}_i 
    \geq f_{\max} \times \max(\lbrace \hat{\importance}_i \rbrace) \nonumber
\end{align}
where $f_{\rm max}$ serves as a threshold relative to the peak value
and $n_{\rm pad}$ pads the starting/ending iteration. In other words,
we compute the importance values $\hat{\importance}_i$ 
over the existing set of samples,
compute the minimum $i^{\rm low}$ and maximum $i^{\rm high}$ iterations
where the importance is above a threshold $f_{\max}$ 
relative to the peak, 
and shift the final values by $n_{\rm pad}$.
By default, {\dynesty} assumes $f^\posterior=0.8$
(80\% posterior vs 20\% evidence), $f_{\max}=0.8$
(80\% thresholding), and $n_{\rm pad} = 1$.

Once we have computed $[i^{\rm low}, i^{\rm high}]$,
we can then just start a new Static Nested Sampling 
run that samples from the
\textit{constrained} prior between 
$[\likelihood^{\rm low}, \likelihood^{\rm high})$.
In the case where $\likelihood^{\rm low}=0$, 
this is just the original prior $\prior(\params)$
and our Static Nested Sampling run is 
identical to Algorithm \ref{alg:static}
except with stopping criteria 
$\likelihood(\params) \geq \likelihood^{\rm high}$.
If $\likelihood^{\rm low}>0$, however,
then we are instead starting \textit{interior} 
to the prior and thus not 
fully integrating over it. So while those new samples will improve
the relative posterior resolution $\Delta \ln X_i$ and thus the posterior
estimate $\hat{\posterior}(\params)$, they will not actually improve
the evidence estimate $\hat{\evidence}$.

Finally, we need to ``merge'' our new set of $N'$ samples 
$\lbrace \params_1', \dots, \params'_{N'} \rbrace$ into our original
set of samples $\lbrace \params_j \rbrace$. This process
is straightforward and can be accomplished following the procedure
outlined in Appendix \ref{ap:nested}.
We are then left with a combined set of samples
$\lbrace \params_1, \dots, \params_{N+N'} \rbrace$ 
with new associated prior volumes 
$\lbrace X_1, \dots, X_{N+N'} \rbrace$
and a variable number of live points 
$\lbrace K_1, \dots, K_{N+N'} \rbrace$
at every iteration.

\subsection{Estimating the Prior Volume}
\label{subsec:dynamic_volume}

As shown in Appendix \ref{ap:nested}, we can reinterpret the
results from \S\ref{subsec:nested_volume} as a consequence
of the two different ways Nested Sampling traverses the prior volume.
In the first case, where the number of live points $K_i \geq K_{i-1}$
increases or stays the same, we know that we have (possibly) added live points
and then \textit{replaced} the one with the lowest likelihood $\likelihood^{\min}$.
In this case, the prior volume experiences 
exponential shrinkage such that
\begin{equation}
    \mean{\Delta \ln \hat{X}_i} 
    = -\frac{1}{K_i}
\end{equation}

In the second case, where the number of live points $K_{j+1} < K_j$ strictly
decreases, we know that we have \textit{removed} the live point(s)
with the lowest likelihood $\likelihood^{\min}$. For each of the
$k$ iterations where this continues to occur, the prior volume experiences
uniform shrinkage such that
\begin{equation}
    \mean{\frac{\Delta \hat{X}_{j+k}}{\hat{X}_j}}
    = \frac{1}{K_j+1}
\end{equation}

In Static Nested Sampling, these two regimes are cleanly divided, with the
main set of dead points traversing the prior volume exponentially
and the final set of ``recycled'' live points traversing it uniformly.
In Dynamic Nested Sampling, however, we are constantly switching between
exponential and uniform shrinkage as we increase or decrease the
number of live points at a given iteration.

\subsection{Stopping Criterion}
\label{subsec:dynamic_stop}

The implementation of Static Nested Sampling outlined in
Algorithm \ref{alg:static} generally \textit{exclusively} 
targets evidence estimation. This gives a
natural stopping criterion (see \S\ref{subsec:nested_stop})
to terminate sampling once we believe that
we have integrated over a majority of the posterior $\posterior(\params)$
such that additional samples will no longer improve our evidence
estimate $\hat{\evidence}$.

In the Dynamic Nested Sampling case, however, we are no longer \textit{just}
interested in computing the evidence. Because we now have the flexibility
to vary the number of live points $K_i$ over time, we are also interested
in the \textit{properties} of our integral (and the samples that
comprise the integrand) in addition to the question of whether 
our integral has converged.

This flexibility necessitates the introduction of more complex
stopping criteria to assess whether those alternative properties
are behaving as expected. Similar to \S\ref{subsec:dynamic_live},
we consider a stopping criteria of the form:
\begin{equation}
    \stopping
    = s^{\posterior} \stopping^{\posterior}
    + (1 - s^{\posterior}) \stopping^{\evidence} < \epsilon
\end{equation}
where $\epsilon$ is our tolerance, $\stopping^{\posterior}$ is the
posterior stopping criterion, $\stopping^{\evidence}$ is the evidence
stopping criterion, and $s^{\posterior}$ is the relative amount of 
weight given to $\stopping^{\posterior}$ over $\stopping^{\evidence}$.

We define our stopping criterion to be the amount of 
\textit{fractional uncertainty} in the current posterior
$\hat{\posterior}(\params)$ and evidence $\hat{\evidence}$ estimates.
For the posterior $\posterior(\params)$, we start by defining
``posterior noise'' to be the Kullback-Leibler (KL) divergence
\begin{align}
    &H(\hat{\posterior}'||\hat{\posterior}) 
    \equiv \meanwrt{\ln \hat{\posterior}' - \ln \hat{\posterior}}
    {\hat{\posterior}'} \\
    &= \int_{\Omega_{\params}} \hat{\posterior}'(\params)
    \ln \hat{\posterior}'(\params) \deriv \params 
    - \int_{\Omega_{\params}} \hat{\posterior}'(\params)
    \ln \hat{\posterior}(\params) \deriv \params 
\end{align}
between the posterior estimate $\hat{\posterior}'(\params)$ from
a random hypothetical Nested Sampling run with the same setup
and our current estimate $\hat{\posterior}(\params)$.
This can be interpreted as the ``information loss'' due
to random noise in our posterior estimate $\hat{\posterior}(\params)$.
Our proposed posterior stopping criteria
is then
\begin{equation}
    \stopping^{\posterior} 
    \equiv \frac{1}{\epsilon^{\posterior}}
    \frac{\stddev{H(\hat{\posterior}'||\hat{\posterior})}}
    {\mean{H(\hat{\posterior}'||\hat{\posterior})}}
\end{equation}
where $\epsilon^{\posterior}$ normalizes the posterior deviation to a desired scale.
For the evidence $\evidence$, this is just the estimated fractional
scatter between the evidence estimates $\hat{\evidence}'$
from random hypothetical Nested Sampling runs with the same setup.
Following \citet{higson+17b}, we opt to compute this in log-space for convenience:
\begin{equation}
    \stopping^{\evidence} 
    \equiv \frac{1}{\epsilon^{\evidence}} \stddev{\ln \hat{\evidence}'}
\end{equation}
where $\epsilon^{\evidence}$ normalizes the evidence deviation to a desired scale.

Unsurprisingly, we do not have access to the distribution
of all hypothetical Nested Sampling runs with the same setup to compute
these exact estimates. However, as with \S\ref{subsec:nested_stop} and
\S\ref{subsec:dynamic_iter}, we \textit{do} have access to noisy estimates of these
quantities via procedures described in \citet{higson+17a} 
and outlined in Appendix \ref{ap:nested} for simulating Nested Sampling errors.
{\dynesty} uses $M$ simulated values of these noisy estimates to
estimate the stopping criteria as:
\begin{align}
    \hat{\stopping}
    &= \frac{s^{\posterior}}{\epsilon^{\posterior}}
    \frac{\stddev{\lbrace \hat{H}_1, \dots, \hat{H}_M \rbrace}}
    {\mean{\lbrace \hat{H}_1, \dots, \hat{H}_M \rbrace}} \nonumber \\
    &+ \frac{(1 - s^{\posterior})}{\epsilon^{\evidence}}
    \stddev{\lbrace \ln \hat{\hat{\evidence}}_1, \dots, \ln \hat{\hat{\evidence}}_M \rbrace}
\end{align}
where the $\hat{\hat{\evidence}}$ notation just emphasizes that we are constructing a noisy
estimator of our already-noisy estimate $\hat{\evidence}$.
By default, {\dynesty} assumes $s^{\posterior}=1$ (100\% focused on reducing
posterior noise), $\epsilon=1$, $\epsilon^\posterior=0.02$, $\epsilon^\evidence=0.1$,
and $M=128$.

\begin{figure*}
	\includegraphics[width=\textwidth]{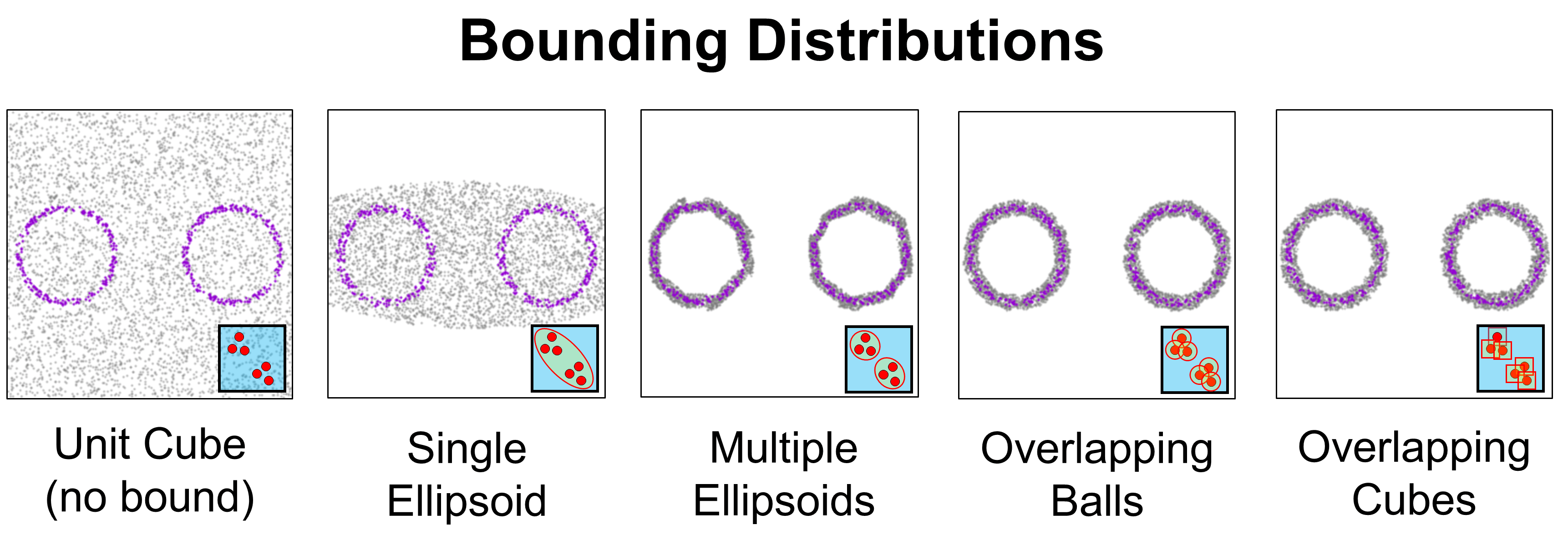}
    \caption{An example highlighting the various bounding distributions
    implemented in {\dynesty}. These include the entire unit cube
    (left), a single ellipsoid (left-middle), multiple overlapping
    ellipsoids (middle), overlapping spheres (right-middle), and overlapping
    cubes (right). The current set of live points
    are shown in purple while draws from the bounding distribution
    are shown in grey. A schematic representation of each bounding
    distribution is shown in the bottom-right-hand corner of each panel.
    See \S\ref{subsec:bounds} for additional details.
    }
    \label{fig:bounding}
\end{figure*}

\section{Implementation}
\label{sec:methods}

Now that we have outlined the basic algorithm and approach behind
Dynamic Nested Sampling, we now turn our attention to the problem
of generating samples from the constrained prior. {\dynesty}
approaches this problem in two parts:
\begin{enumerate}
    \item constructing appropriate \textit{bounding distributions} that encompass
    the remaining prior volume over multiple possible modes and
    \item proposing new live points by 
    generating samples \textit{conditioned on} these bounds.
\end{enumerate}

{\dynesty} contains several options for both constructing bounds
and sampling conditioned on them. We provide an broad overview of each
of these in turn.

\subsection{Bounding Distributions}
\label{subsec:bounds}

In general, {\dynesty} tries to use 
the distribution of the current set of live points 
to try and get a rough idea of the shape and size of the various
regions of prior volume that we are currently sampling. These are
then used to condition various sampling methods to try and improve
the efficiency. There are five bounding methods currently implemented in
\dynesty:
\begin{itemize}
    \item no bounds (i.e. the unit cube),
    \item a single ellipsoid,
    \item multiple ellipsoids,
    \item many overlapping balls, and
    \item many overlapping cubes.
\end{itemize}

In general, single ellipsoids tend to perform reasonably well
at estimating structure when the likelihood is roughly Gaussian
and uni-modal. In more complex cases, however, decomposing the live
points into separate clusters with their own bounding ellipsoids
works reasonably well at locating and tracking structure.
In low ($D \lesssim 5$) dimensions,
allowing the live points themselves to define emergent 
structure through many overlapping balls or cubes can perform
better provided the $\likelihood(\params)$ spans similar scales
in each of the parameters. Finally, using no bounds at all
is only recommended as an option of last resort and is mostly
relevant when performing systematics checks or if the number
of live points $K \ll D^2/2$ is small relative
to the number of possible parameter covariances.

In addition to these various options, {\dynesty} also tries to increase
the volume of all bounds by a factor $\alpha$
to be conservative about the size of the constrained prior.
While this is generally assumed to take a constant value of $\alpha=1.25$,
it can also be derived ``on the fly'' using bootstrapping methods following
the approach outlined in \citet{buchner16}. Deriving accurate volume
expansion factors are extremely important when 
sampling uniformly but are less relevant for other sampling schemes
that are more robust to the exact sizes of the bounds (see \S\ref{subsec:samples}).

By default, {\dynesty} uses multiple ellipsoids to construct the bounding distribution.
A summary of the various bounding methods can be found in 
Figure \ref{fig:bounding}. We describe these each in turn below.

\subsubsection{Unit Cube}
\label{subsubsec:unit}

The simplest case of using the entire unit cube
(i.e. simple rejection sampling over the entire
prior $\prior(\params)$ with no limits)
can be useful in a few edge cases
where the number of live points $K$
is small compared to the number of dimensions $D$,
or where users are interested in performing tests
to verify sampling behavior.

\subsubsection{Single Ellipsoid}
\label{subsubsec:single}

As shown in \citep{mukherjee+06},
a single bounding ellipsoid can be effective if the posterior
is unimodal and roughly Gaussian. {\dynesty} uses a scaled version
of the empirical covariance matrix $\cov'=\gamma\cov$
centered on the empirical mean $\boldsymbol{\mu}$ of the current
set of live points to determine the size
and shape of the ellipsoid, where $\gamma$ is set so
the ellipsoid encompasses all available live points.

\subsubsection{Multiple Ellipsoids}
\label{subsubsec:multi}

By default, {\dynesty} does not assume the posterior is unimodal or
Gaussian and instead tries to bound the live points using
a set of (possibly overlapping) ellipsoids. These are constructed
using an iterative clustering scheme following the algorithm outlined
in \citet{shaw+07} and \citet{ferozhobson08} 
and implemented in the online package
\href{http://kylebarbary.com/nestle/}{\nestle}.\footnote{
{\dynesty} is built off of {\nestle} with the permission of
its developer Kyle Barbary.}
In brief, we start by constructing a bounding ellipsoid
over the entire collection of live points. We then initialize
2 $k$-means clusters at the endpoints of the major axes, optimize
their positions, assign live points to each cluster, and
construct a new pair of bounding ellipsoids for each new cluster
of live points. The decomposition is accepted if the combined volume of the
subsequent pair of ellipsoids is substantially smaller. This process
is then performed recursively until no decomposition is accepted.

By default, {\dynesty} tries to be substantially more conservative
when decomposing live points into separate clusters and bounding ellipsoids
than alternative approaches used in \texttt{MultiNest} \citep{ferozhobson08,feroz+13}.
This algorithmic choice, which can substantially reduce 
the overall sampling efficiency, is made in order to avoid
``shredding'' the posterior into many tiny islands 
of isolated live point clusters. As shown in \citet{buchner16},
that behavior can lead to biases in the estimated evidence
$\hat{\evidence}$ and posterior $\hat{\posterior}(\params)$.


\subsubsection{Overlapping Balls}
\label{subsubsec:balls}

An alternate approach to using bounding ellipsoids is to 
allow the current set of live points themselves to define 
emergent structure. The simplest approach used in {\dynesty} follows 
\citet{buchner16,buchner17} by assigning a $D$-dimensional ball (sphere) 
with radius $r$ to each live point, where $r$
is set using bootstrapping and/or leave-one-out
techniques to encompass $\geq 1$ other live points.
One benefit to this approach over using multiple ellipsoids
(which can depend sensitively on the clustering schemes)
is that it is almost entirely free of tuning parameters,
with the overall behavior only
weakly dependent on the number of bootstrap realizations.

\subsubsection{Overlapping Cubes}
\label{subsubsec:cubes}

As with the set of overlapping balls, {\dynesty} also 
implements a similar algorithm based on \citet{buchner16,buchner17} 
involving overlapping cubes with half-side-length $\ell$. 
As \S\ref{subsubsec:balls}, $\ell$ is derived using 
either bootstrapping and/or leave-one-out techniques
so that the cubes encompass $\geq 1$ other live points.

\subsection{Sampling Methods}
\label{subsec:samples}

\begin{figure*}
	\includegraphics[width=0.75\textwidth]{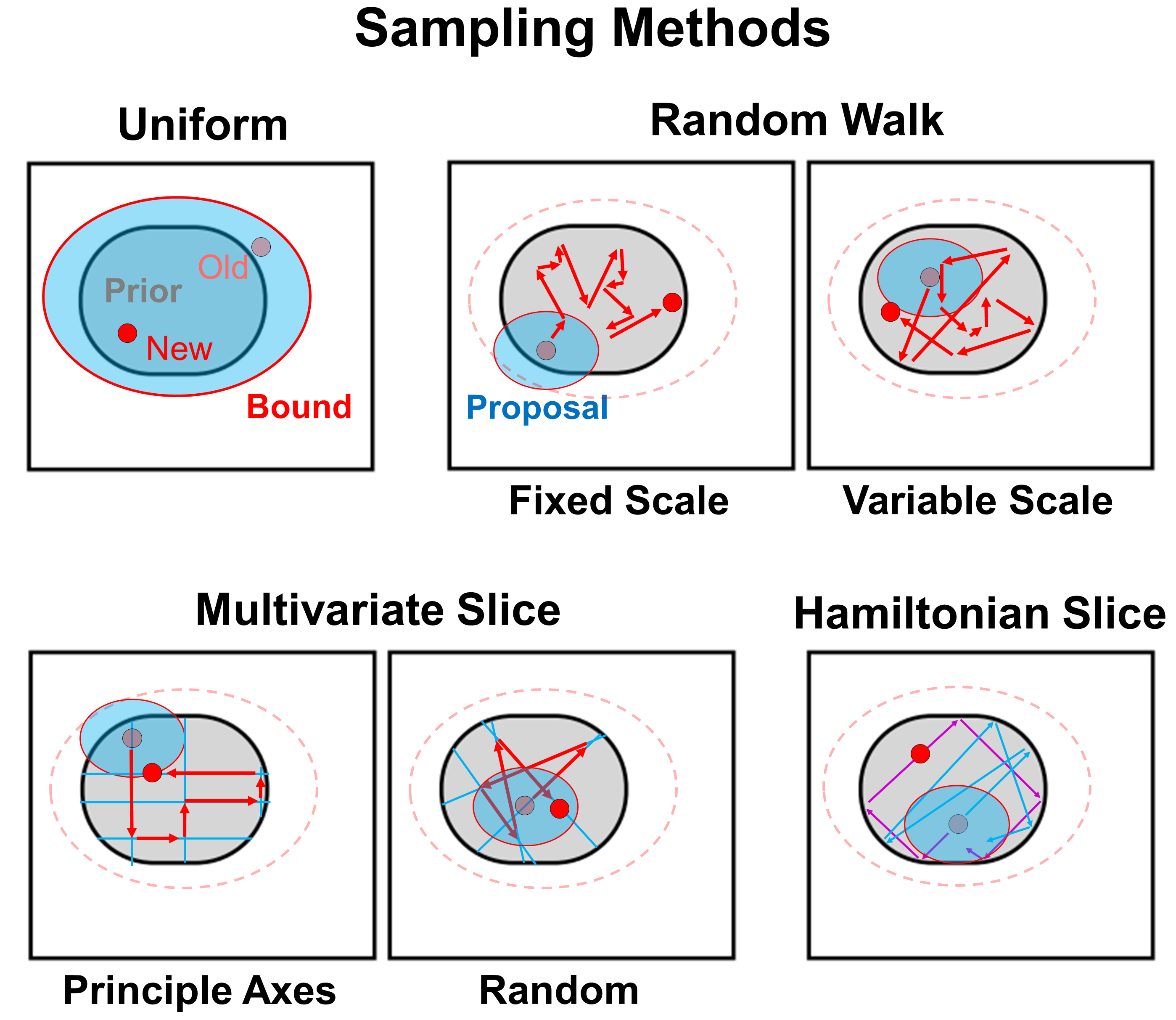}
    \caption{A schematic illustration of the different sampling
    methods implemented in {\dynesty}. These include:
    uniform sampling from the bounding distribution (top-left),
    random walks proposals starting from a random live point based
    on the bounding distribution (top-right) with either fixed or
    variable scale-lengths for proposals, multivariate slice sampling
    proposals starting from a random live point (bottom-left) using
    either the principle axes or a random direction sampled from
    the bounding distribution, and Hamiltonian slice sampling
    away from a random live point forwards and backwards in time
    (bottom-right). See \S\ref{subsec:samples} for additional details.}
    \label{fig:sampling}
\end{figure*}

Once a bounding distribution has been constructed,
{\dynesty} generates samples \textit{conditioned on} those bounds.
In general, this follows a strategy of 
\begin{equation}
    f(s\cov_b, \params) \rightarrow \params'
\end{equation}
where $\cov_b$ is the covariance associated with a particular
bound $b$ (e.g., an ellipsoid), $\params$ is
the starting position, $\params '$ is the final proposed position,
and $s \sim 1$ is a scale-factor that
is adaptively tuned over the course of a run to ensure
optimal acceptance rates.

{\dynesty} implements four main approaches to generating samples:
\begin{itemize}
    \item uniform sampling,
    \item random walks,
    \item multivariate slice sampling, and
    \item Hamiltonian slice sampling.
\end{itemize}
These each are designed for different regimes. Uniform sampling
can be relatively efficient in lower dimensions where
the bounding distribution can approximate the prior volume
better but struggles in higher dimensions since it
is extremely sensitive to the size of the bounds. Random
walks are less sensitive to the size of the bounding distribution
and so tend to work better than uniform sampling
in moderate dimensional spaces but still struggle in high-dimensional
spaces because of the exponentially increasing amount of volume it
needs to explore. Multivariate and Hamiltonain slice sampling
often performs better in these high-dimensional regimes by
avoiding sampling directly from the volume and
taking advantage of gradients, respectively.

In addition to each method's performance in various regimes,
there is also a fundamental qualitative difference between uniform sampling
and the other sampling approaches outlined above. Uniform sampling, by
construction, can only sample directly from the bounding distribution. This
makes it uniquely sensitive to the assumption that the bounds entirely
encompass the current prior volume at a given iteration, which is never
fully guaranteed \citep{buchner16}.
By contrast, the other sampling methods are MCMC-based: they generate
samples by ``evolving'' a current live point to a new position. This
allows them to generate samples outside the bounding distribution, making
them less sensitive to this assumption.

By default, {\dynesty} resorts to uniform sampling when
the number of dimensions $D < 10$, 
random walks when $10 \leq D \leq 20$,
and Hamiltonian/multivariate slice sampling when $D > 20$ 
if a gradient is/is not provided.
A summary of the various sampling methods can be found in 
Figure \ref{fig:sampling}. We describe these each in turn below.

\subsubsection{Uniform Sampling}
\label{subsubsec:unif}

If we assume that our bounding distribution $B(\params)$
encloses the constrained prior $\prior_\lambda(\params)$,
the most direct approach to generating samples
from the bounds is to sample from them uniformly.
This procedure by construction produces entirely independent
samples between each iteration $i$, and tends to work best when the volume
of the bounds $X_B(\lambda)$ is roughly the same order of magnitude
as the current prior volume $X(\lambda)$ (leading to $\gtrsim 10\%$
acceptance rates).

In general, the procedure for generating uniform samples
from overlapping bounds is straightforward 
\citep[see, e.g.,][]{ferozhobson08,buchner16}:
\begin{enumerate}
    \item Pick a bound $b$ at random with probability
    $p_b \propto X_b$ proportional to its volume $X_b$.
    \item Sample a point $\params_b$ uniformly
    from the bound.
    \item Accept the point with probability $1/q$,
    where $q \geq 1$ is the number of bounds $\params_b$
    lies within.
\end{enumerate}
This approach ensures that any proposed sample will be drawn from
the bounding distributing $B(\params)$ comprised of the \textit{union}
of all bounds, which has a volume $X_B \leq \sum_{b=1}^{N_b} X_b$
that is strictly less than or equal to the sum of the volumes of each
individual bound. 

Generating samples uniformly from the bounds in \S\ref{subsec:bounds}
falls into two cases: cubes and ellipsoids.
Generating points from an $D$-cube centered at $\params_b$ 
with half-side-length $\ell$ is trivial and can be accomplished via:
\begin{enumerate}
    \item Generate $D$ iid uniform random numbers 
    $\mathbf{U}=\lbrace U_1, \dots, U_D \rbrace$ from $[-\ell, \ell]$.
    \item Set $\params' = \mathbf{U} + \params_b$.
\end{enumerate}

Generating points from an ellipsoid centered at $\params_b$
with covariance $\cov_b$ with matrix square-root
$\cov_b^{1/2}$ is also straightforward but slightly more involved:
\begin{enumerate}
    \item Generate $D$ iid standard normal random numbers
    $\mathbf{Z} = \lbrace Z_1, \dots Z_D \rbrace$.
    \item Compute the normalized vector
    $\mathbf{V} \equiv \mathbf{Z}/||\mathbf{Z}||$.
    \item Draw a standard uniform random number $U$
    and compute $\mathbf{S} \equiv U^D \mathbf{V}$.
    \item Set $\params = \cov_b^{1/2} \mathbf{S} + \params_b$.
\end{enumerate}
Step (ii) creates a random vector $\mathbf{V}$ that is uniformly 
distributed on the \textit{surface} of the $D$-sphere. Step (iii)
randomly moves $\mathbf{V}\rightarrow\mathbf{S}$
to an interior radius $r\in(0,1)$ based on the fact that the volume of a 
$D$-sphere scales as $V(r) \propto r^D$. Finally,
step (iv) adjusts the scale, shape, and center to match that
of the bounding ellipsoid.

\subsubsection{Random Walks}
\label{subsubsec:rwalk}

An alternative approach to sampling uniformly
within the bounding distribution $B(\params)$ 
is to instead to try and propose new positions by ``evolving''
a given live point $\params_k \rightarrow \params'$ to a new
position. Since $\likelihood(\params_k) \geq \likelihood^{\min}_i$
at a given iteration by definition, this procedure also 
guarantees that we will be
generating samples exclusively within the constrained prior
$\prior_\lambda(\params)$. 

One straightforward approach to ``evolving'' a live point
to a new position is to consider sampling from the
constrained prior using a simple Metroplis-Hastings
\citep[MH;][]{metropolis+53,hastings70}
MCMC algorithm:
\begin{enumerate}
    \item Propose a new position $\params' \sim Q(\params|\params_k)$
    from the proposal distribution $Q(\params|\params_k)$ 
    starting from $\params_k$.
    \item Move to $\params'$ with probability
    $A=\frac{\prior_\lambda(\params')}{\prior_\lambda(\params_k)}
    \frac{Q(\params_k|\params')}{Q(\params'|\params_k)}$. Otherwise,
    stay at $\params_k$.
    \item Repeat (i)-(ii) for $N_{\rm walks}$ iterations.
\end{enumerate}
Since the constrained prior is flat (see \S\ref{subsec:nested_samples}),
the ratio of the constrained prior
values is by definition $1$. Likewise, if we choose a symmetric
proposal distribution $Q(\params|\params_k)$, then the ratio of the
proposal distributions also evaluates to $1$. This procedure then reduces
to simply accepting a new point if it is within the constrained prior
with $\likelihood(\params_i) \geq \lambda$ and rejecting it otherwise.
By default, {\dynesty} takes $N_{\rm walks} = 25$.

{\dynesty} implements two forms of the proposal $Q(\params|\params_k)$.
The default option is to propose new positions 
uniformly from an associated ellipsoid
centered on $\params_k$ with covariance $\cov_b$, where $\cov_b$ is one
of the bounding distributions that encompasses $\params_k$ (selected randomly).
The second follows the same form as the first, 
except the covariance $\cov_b$ is re-scaled
at each subsequent proposal $t \leq N_{\rm walks}$ by $\gamma$ following
the procedure outlined in \citet{siviaskilling06}:
\begin{equation}
    \alpha(t) 
    = \begin{cases}
    e^{1/N_{\rm acc}(t)} \times \gamma(t-1) & \frac{N_{\rm acc}(t)}{t} > f_{\rm acc} \\
    e^{-1/N_{\rm rej}(t)} \times \gamma(t-1) & \frac{N_{\rm acc}(t)}{t} < f_{\rm acc} \\
    \gamma(t-1) & \frac{N_{\rm acc}}{t} = f_{\rm acc}
    \end{cases}
\end{equation}
where $N_{\rm acc}(t)$ and $N_{\rm rej}(t)$ is the total 
number of accepted and rejected
proposals by iteration $t$, respectively, 
$f_{\rm acc}$ is the desired acceptance fraction,
and $\gamma(t=0)=1$. By default, {\dynesty} targets $f_{\rm acc} = 0.5$.

\subsubsection{Multivariate Slice Sampling}
\label{subsubsec:slice}

In higher dimensions, rejection sampling-based methods such as the random
walk proposals outlined in \S\ref{subsubsec:rwalk} can become progressively
more inefficient. To remedy this, {\dynesty} includes slice sampling
\citep{neal03} routines designed to sample from the constrained prior 
$\prior_\lambda(\params)$. These are based on the ``stepping out''
method proposed in \citet{neal03} and \citet{jasaxiang12},
which works as follows in the single-variable
case starting from the position $x_k$ of the $k$th live point:
\begin{enumerate}
    \item Draw a standard uniform random number $U$.
    \item Set the left bound $L = x_k - w U$ and the
    right as $R = L + w$ where $w$ is the starting ``window''.
    \item While $\likelihood(L) \geq \lambda$, extend
    the position of the left bound $L$ by $w$. Repeat this procedure
    for $R$.
    \item Sample a point $x' \sim \Unif(L, R)$ uniformly on the
    interval from $L$ to $R$.
    \item If $\likelihood(x') > \lambda$, accept $x'$. Otherwise,
    reassign the corresponding bound to be $x'$ 
    ($L$ if $x' < x$ and $R$ otherwise) and repeat steps (iv)-(v).
\end{enumerate}

When sampling in higher dimensions, the single-variable update
outlined above can be interpreted as a Gibbs sampling update
\citep{gemangeman87} where instead of drawing $\params$ directly
we instead update each component in turn
\begin{equation}
    \params' \sim \prior_\lambda(\params)
    \Rightarrow
    \begin{cases}
    \Theta_1 \sim \prior_\lambda(\Theta_1|\params_{\setminus 1}) \\
    \quad\quad\quad\vdots \\
    \Theta_D \sim \prior_\lambda(\Theta_D|\params_{\setminus D})
    \end{cases}
\end{equation}
where $\params_{\setminus i}$ are the set 
of $D-1$ parameters excluding $\Theta_i$.
We then repeat this procedure for $N_{\rm slices}$ iterations.
By default {\dynesty} takes $N_{\rm slices} = 5$.

\begin{figure*}
	\includegraphics[width=\textwidth]{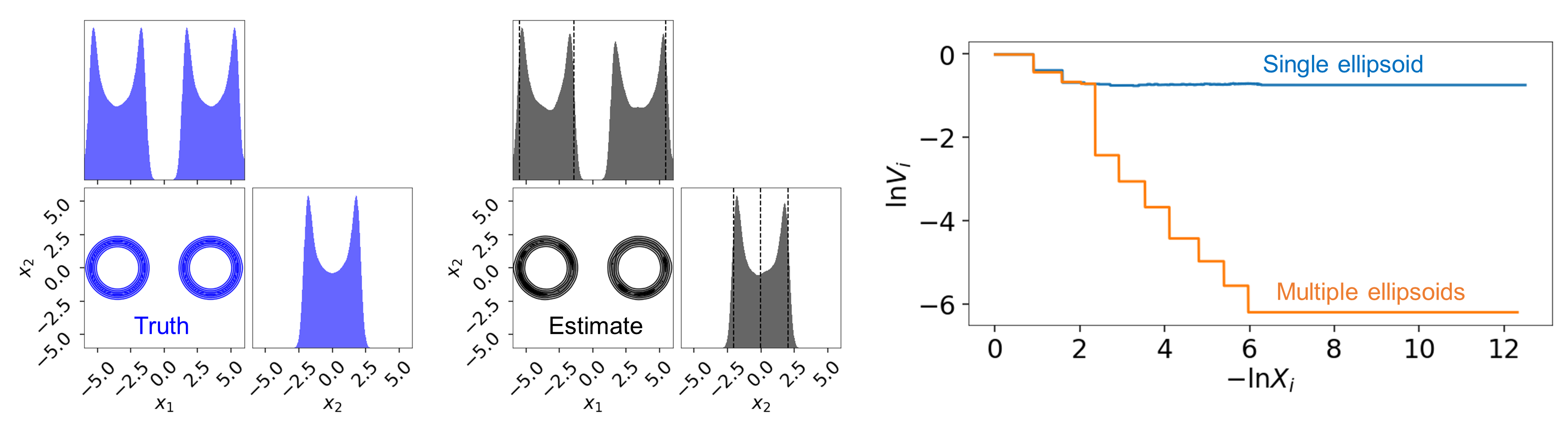}
    \caption{Illustration of {\dynesty}'s performance using
    multiple bounding ellipsoids and uniform sampling over
    2-D Gaussian shells (highlighted 
    in Figure \ref{fig:bounding}) meant to test the code's bounding
    distributions. 
    \textit{Left}: A smoothed corner plot showing the
    exact 1-D and 2-D marginalized posteriors of
    the target distribution. 
    \textit{Middle}: As before, but now showing
    the final distribution of weighted samples. 
    \textit{Right:} The volume of the bounding distribution
    when using a single ellipsoid (blue) versus
    multiple ellipsoids (orange) over the course of the run.
    Since a single ellipsoid is a poor model for this distribution,
    its volume quickly saturates as it becomes unable to accurately
    capture the distribution of live points. Allowing the
    bounding distribution to be modeled by multiple ellipsoids
    allows for {\dynesty} to capture the more complex structure
    as the live points move increasingly into organized rings.
    }
    \label{fig:shells}
\end{figure*}

This procedure is generally robust, although it can
introduce longer correlation times if there are strong covariances
between parameters. To mitigate this, {\dynesty} by default
executes single-variable slice sampling updates along the principle axes
$\mathbf{V}_b \equiv \lbrace \mathbf{v}_{1,b}, \dots, \mathbf{v}_{D,b} \rbrace$
associated with the covariance $\cov_b$ from a given bound $b$. This
allows us to automatically set both the direction $\mathbf{v}_{i,b}$
and associated scale $||\mathbf{v}_{i,b}||$ of the window while trying to
reduce the correlations among sets of parameters.

Alternately, instead of executing a full Gibbs update by rotating through
the entire set of parameters in turn, we can sample along a random trajectory
$\mathbf{v}'$ through the prior instead. This procedure is similar to that
implemented in {\polychord} \citep{handley+15},
except that rather than ``whitening'' the set of live points using
the associated $\cov_b$ we instead draw $\mathbf{v}'$ from the \textit{surface}
of the corresponding bound with covariance $\cov_b$. Provided a suitable
number of $N_{\rm slices} \sim D$, this procedure also can generate suitably independent
new positions $\params'$.

\subsubsection{Hamiltonian Slice Sampling}
\label{subsubsec:hslice}

Over the past two decades, sampling methods have increasingly
attempted to incorporate gradients to improve their overall
performance, especially in high-dimensional spaces. The most common
class of methods are based on Hamiltonian Monte Carlo 
\citep[HMC;][]{neal12,betancourt+17},
whereby a particle at a given position $\mathbf{x}$
is assigned a mass matrix $\mathbf{M}$
and some momentum $\mathbf{p}$ and allowed to sample from the
joint distribution
\begin{equation}
    P(\mathbf{x}, \mathbf{p}|\mathbf{M}) 
    \propto \exp\left[-\mathcal{H}(\mathbf{x}, \mathbf{p} | \mathbf{M})\right]
\end{equation}
where
\begin{equation}
    \mathcal{H}(\mathbf{x}, \mathbf{p} | \mathbf{M})
    \equiv U(\mathbf{x}) + K(\mathbf{p}|\mathbf{M})
    \equiv -\ln\left[\prior(\mathbf{x})\likelihood(\mathbf{x})\right]
    + \frac{1}{2} \mathbf{p}^T \mathbf{M}^{-1} \mathbf{p}
\end{equation}
is the Hamiltonian of the system with a ``potential energy'' $U(\mathbf{x})$ and
``kinetic energy'' $K(\mathbf{p}|\mathbf{M})$, and $T$ is the transpose operator.
Typically, proposals are generated by sampling the
momentum from the corresponding multivariate
Normal (Gaussian) distribution
\begin{equation}
    \mathbf{p} \sim \Normal{\mathbf{0}}{\mathbf{M}} \:,
\end{equation}
with mean $\mathbf{0}$ and covariance $\mathbf{M}$,
evolving the system via Hamilton's equations from
$\mathcal{H}(\mathbf{x}, \mathbf{p}) \rightarrow
\mathcal{H}(\mathbf{x}', \mathbf{p}')$, and then accepting the
new position based on the MH acceptance criteria outlined in
\S\ref{subsubsec:rwalk}. In other words, at
each iteration we randomly assign a given particle
some mass and velocity and then have it explore
the potential defined by the (log-)posterior.

\begin{figure*}
	\includegraphics[width=0.9\textwidth]{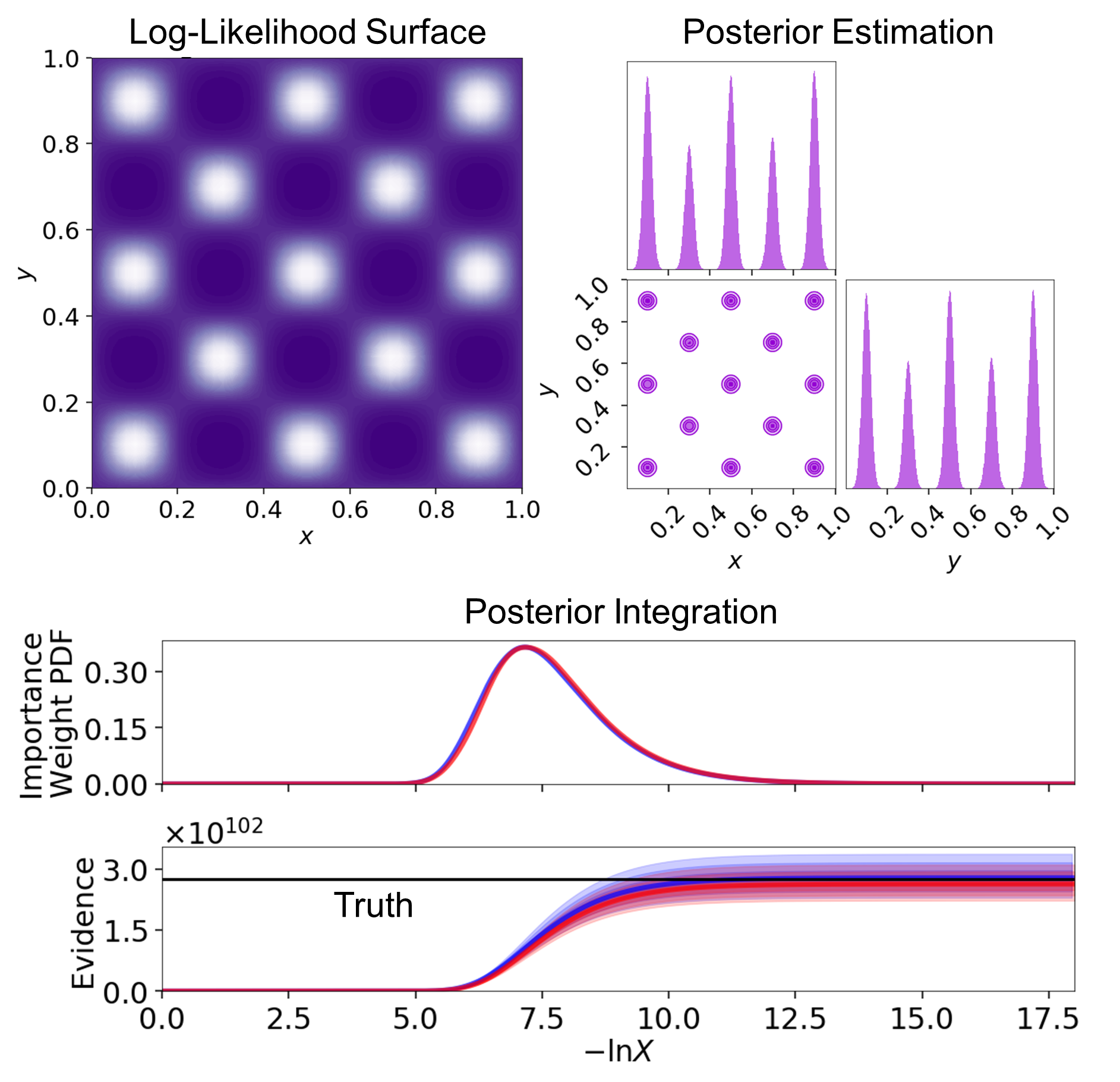}
    \caption{Illustration of {\dynesty}'s performance using
    multiple bounding ellipsoids and overlapping balls
    with uniform sampling over the 2-D ``Eggbox''
    distribution meant to test the code's bounding
    distributions. 
    \textit{Top left}: The true log-likelihood surface 
    of the Eggbox distribution. 
    \textit{Top right}: A smoothed corner plot showing the
    1-D and 2-D marginalized posteriors of the final
    distribution of weighted samples from a posterior-oriented
    Dynamic Nested Samplig run. 
    \textit{Bottom}: The importance weight PDF $p(X)$ (top)
    and corresponding evidence estimate $\hat{\evidence}$
    with 1, 2, and 3-sigma uncertainties (bottom) from two independent
    evidence-oriented Dynamic Nested Sampling runs using
    multiple ellipsoids (blue) and overlapping balls (red) as bounding
    distributions.
    }
    \label{fig:eggbox}
\end{figure*}

As with the previous methods, this approach simplifies dramatically
when sampling over the constrained prior $\pi_\lambda(\params)$.
In that case, since the distribution is flat, the momentum remains unchanged
until the particle hits the hard likelihood boundary, at which point it
reflects so that
\begin{equation}
    \mathbf{p}' 
    = \mathbf{p} - 2\mathbf{h}
    \frac{\mathbf{p} \cdot \mathbf{h}}{||\mathbf{h}||^2}
\end{equation}
where $\mathbf{h}$ is the gradient at the point of reflection.
This version of the algorithm is referred to
elsewhere as Galilean Monte Carlo \citep{skilling+12,ferozskilling13}
or reflective slice sampling \citep{neal03}.

In practice, since we have to evolve the system discretely, 
there are a few additional caveats to consider. Most importantly, the use of
discrete time-steps means reflection will not occur right \textit{at} the
boundary of the constrained prior but slightly \textit{beyond} it, which does not
guarantee reflections will end up back inside the constrained prior. This
behavior, which arises from larger time-steps,
``terminates'' the particle's trajectory in that particular direction and
leads to inefficient sampling that isn't able to explore the full parameter
space. 

On the other hand, using extremely small time-steps means spending
the vast majority of time evaluating positions along a straight line, which
is also non-optimal. {\dynesty} by default attempts to compromise
between these two behaviors by optimizing the time-step so that
$f_{\rm move} \sim 0.9$ of total steps are spent moving forward passively
instead of reflecting or terminating. In addition, {\dynesty}
by default caps the total number of time-steps to $N_{\rm move} = 100$ 
to prevent trajectories from being evolved indefinitely.

Similar to algorithms such as the No U-Turn Sampler \citep[NUTS;][]{hoffman+11},
{\dynesty} also considers trajectories 
evolved forwards and backwards in time to broaden
the range of possible positions explored in a given proposal. While these 
roughly double the number of overall time-steps, they substantially improve
overall behavior by exploring larger regions of the constrained prior.

{\dynesty} employs two additional schemes to try and further
mitigate discretization effects on the sampling procedure described above.
First, the time-step used at a given iteration is allowed to vary
randomly by up to $30\%$ following recommendations from \citet{neal12}.
This helps to suppress resonant behavior that can arise from poor choices of
time-steps without substantially impacting overall performance. Second,
rather than merely accepting positions at the end of a trajectory,
{\dynesty} instead tries to sample uniformly from the \textit{entire trajectory}
by treating it as a set of slices defined by 
$(\params_L^i, \params^i, \params_R^i)$ left-inner-right position tuples.
New samples are then proposed via the following scheme:
\begin{enumerate}
    \item Compute the length $\ell_i$ of each line segment $(\params_L^i, \params_R^i)$.
    \item Selecting a line segment $i$ at random proportional to its length.
    \item Sample a point $\params'$ uniformly on the
    line segment defined by $(\params_L^i, \params_R^i)$.
    \item If $\likelihood(\params') > \lambda$, accept $\params'$. Otherwise,
    reassign the corresponding bound to be $\params'$ ($\params_L^i$ if 
    $\params'$ is on the line segment $[\params_L^i,\params^i)$ and $\params_R^i$
    otherwise) and repeat steps (i)-(iv).
\end{enumerate}

While there are a variety of possible approaches to 
applying HMC-like methods to Nested Sampling other than the basic procedure
outlined above, we defer any detailed comparisons between them
to possible future work.

\section{Tests}
\label{sec:tests}

Here, we examine {\dynesty}'s performance on a variety
of toy problems designed to stress-test various aspects of the code.
Additional tests can also be found
\href{https://dynesty.readthedocs.io/en/latest/examples.html}{online}.

\begin{figure*}
	\includegraphics[width=\textwidth]{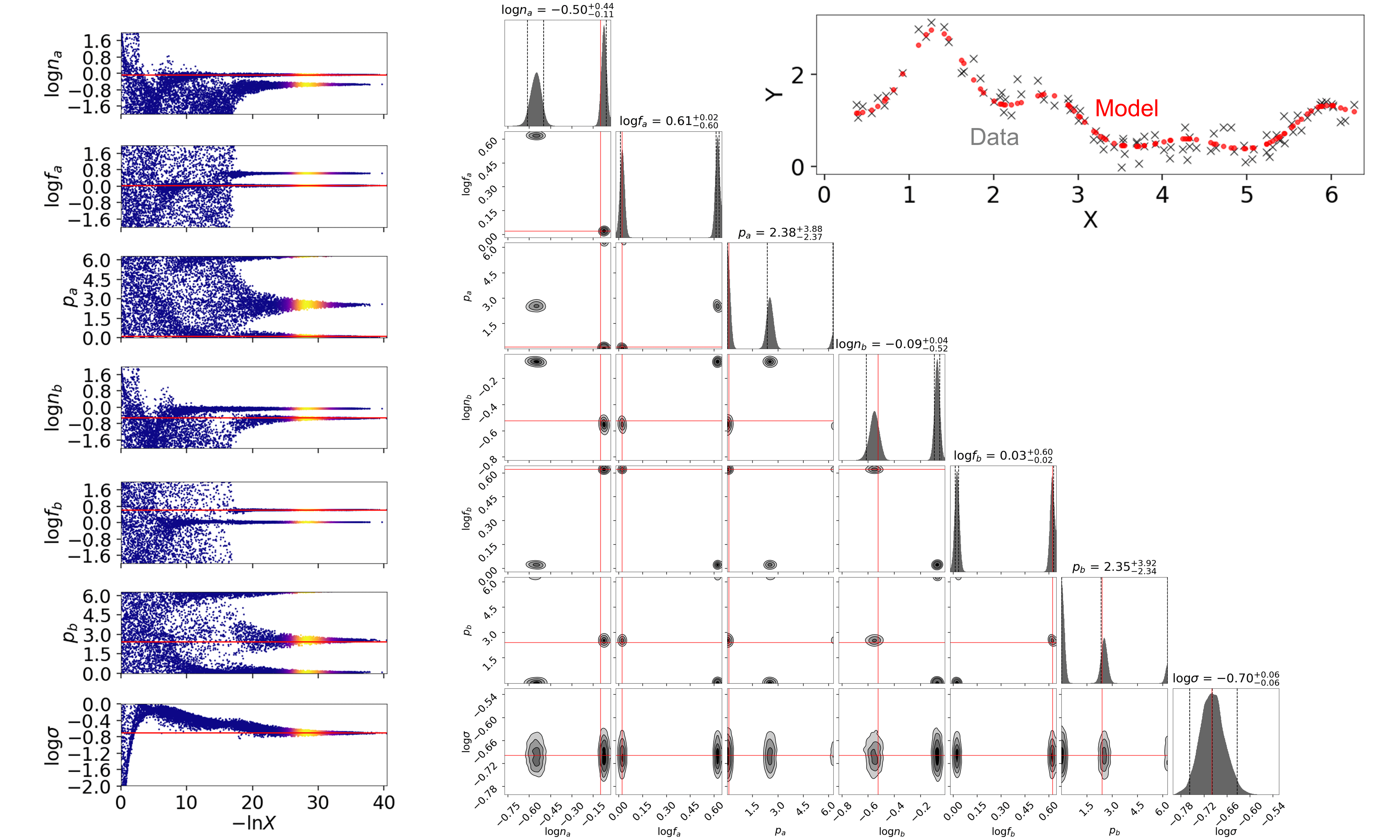}
    \caption{Illustration of {\dynesty}'s performance
    using multiple bounding ellipsoids and multivariate slice sampling
    over principle axes to model an ``Exponential Wave'' signal 
    meant to test the code's bounding
    distributions and incorporation of
    periodic boundary conditions. 
    \textit{Left}: Trace plots showing the 1-D positions
    of samples (dead points) over the course of the run,
    colored by their estimated importance weight PDF $p(X)$.
    The true model parameters are shown highlighted in red.
    We see that even though the underlying structure
    of the distribution spans many different scales and 
    emerges in different stages,
    {\dynesty} is able to confidently identify the final
    two modes and converge to the underlying model parameters. 
    \textit{Middle}: A corner plot showing the 1-D and 2-D marginalized
    posteriors from the distribution of the final weighted samples.
    The true model parameter values are shown in red. The 2.5\%, 50\%,
    and 97.5\% percentiles (i.e. the 2-sigma credible region) are
    shown as vertical dashed lines. 
    \textit{Top right}: The noisy data (gray crosses)
    and underlying model (red points).}
    \label{fig:expwave}
\end{figure*}


\subsection{Gaussian Shells}
\label{subsec:gaussian_shells}

One standard problem that tests the efficiency of the
ability of bounding distributions
to transition between a flat surface to 
separated, elongated structures is the $D$-dimensional 
``Gaussian shells'' from \citet{ferozhobson08}.
The likelihood of the distribution is defined as
\begin{equation}
    \likelihood(\params) 
    = \textrm{circ}(\params|\mathbf{c}_1, r_1, w_1)
    + \textrm{circ}(\params|\mathbf{c}_2, r_2, w_2)
\end{equation}
where
\begin{equation}
    \textrm{circ}(\params | \mathbf{c}, r, w) 
    = \frac{1}{\sqrt{2\pi w^2}} 
    \exp\left[-\frac{1}{2}\frac{\left(||\params - \mathbf{c}||
    - r\right)^2}{w^2}\right]
\end{equation}
Following \citet{feroz+13}, 
we take the centers $\mathbf{c}_1$ and $\mathbf{c}_2$
of the two positions to be $-3.5$ and $3.5$
in the first dimension and $0$ in all others, respectively,
the radius $r=2$, and the width $w=0.1$. Our prior is
defined to be uniform from $[-6, 6]$ to encompass the majority
of the likelihood and ensure a smooth transition between the
uni-modal starting distribution and the multi-modal target
distribution.

We illustrate {\dynesty}'s performance in the 2-D case
in Figure \ref{fig:shells}. The default configuration
options in {\dynesty} (multiple ellipsoid bounds with
uniform sampling) lead to a roughly $10\%$ sampling
efficiency over the course of $\sim 20$k iterations
when using Dynamic Nested Sampling and lead to excellent
posterior estimates. We also see that the multi-ellipsoidal
decomposition algorithm works as expected, with the 
total volume of the bounding distribution decreasing dramatically
as the live points begin to organize themselves within the
two shells.

\begin{figure*}
	\includegraphics[width=\textwidth]{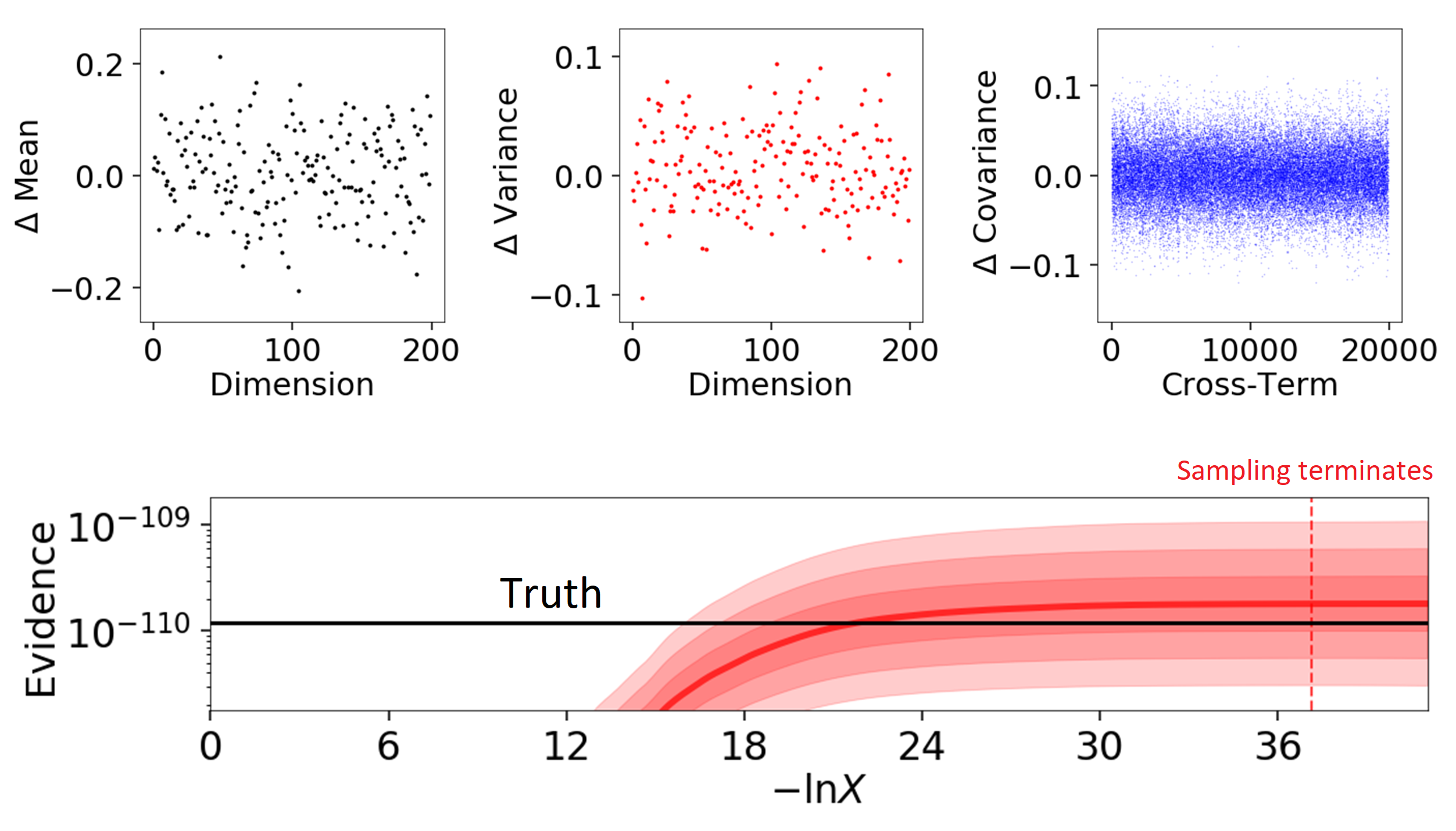}
    \caption{Illustration of {\dynesty}'s performance
    sampling from a 200-D Gaussian using
    Hamiltonian Slice Sampling (\S\ref{subsubsec:hslice})
    with gradients and no bounding distribution with 
    only $K=50$ live points. 
    \textit{Top}: Offsets in the recovered mean (left, black),
    variance (center, red), and covariance cross-terms
    (right, blue) relative to an expected mean of
    $\boldsymbol{\mu} = \mathbf{0}$ and covariance of
    $\cov = (1/2) \, \mathbf{I}$. 
    \textit{Bottom}: The estimated evidence $\hat{\evidence}$
    (red line) along with the 1, 2, and 3-sigma errors (shaded).
    The true value is shown in black, along with the location where
    sampling terminates (dotted red vertical line).}
    \label{fig:200d}
\end{figure*}

\subsection{Eggbox}
\label{subsec:eggbox}

Another distribution we consider to test the ability
of {\dynesty} to track and evolve multiple modes is
the 2-D ``Eggbox'' likelihood from \citet{ferozhobson08},
which we defined as
\begin{equation}
    \likelihood(x, y) 
    = \exp\left\{\left[2 + \cos\left(\frac{5\pi(x-1)}{2}\right)
    \sin\left(\frac{5\pi(y-1)}{2}\right)\right]^5\right\}
\end{equation}
This distribution is periodic over the 2-D unit cube,
with 13 localized modes contained within a given period.
We take our prior to be standard uniform in $x$ and $y$
to limit sampling to one period.

The resulting posterior and evidence estimates from
several posterior-oriented and evidence-oriented Dynamic
Nested Sampling runs are shown in Figure \ref{fig:eggbox}.
{\dynesty} is able to sample from this distribution
quite effectively, with average sampling efficiencies
ranging from $20-40\%$ when sampling uniformly
from the multiple ellipsoids or overlapping balls.

\begin{figure*}
	\includegraphics[width=\textwidth]{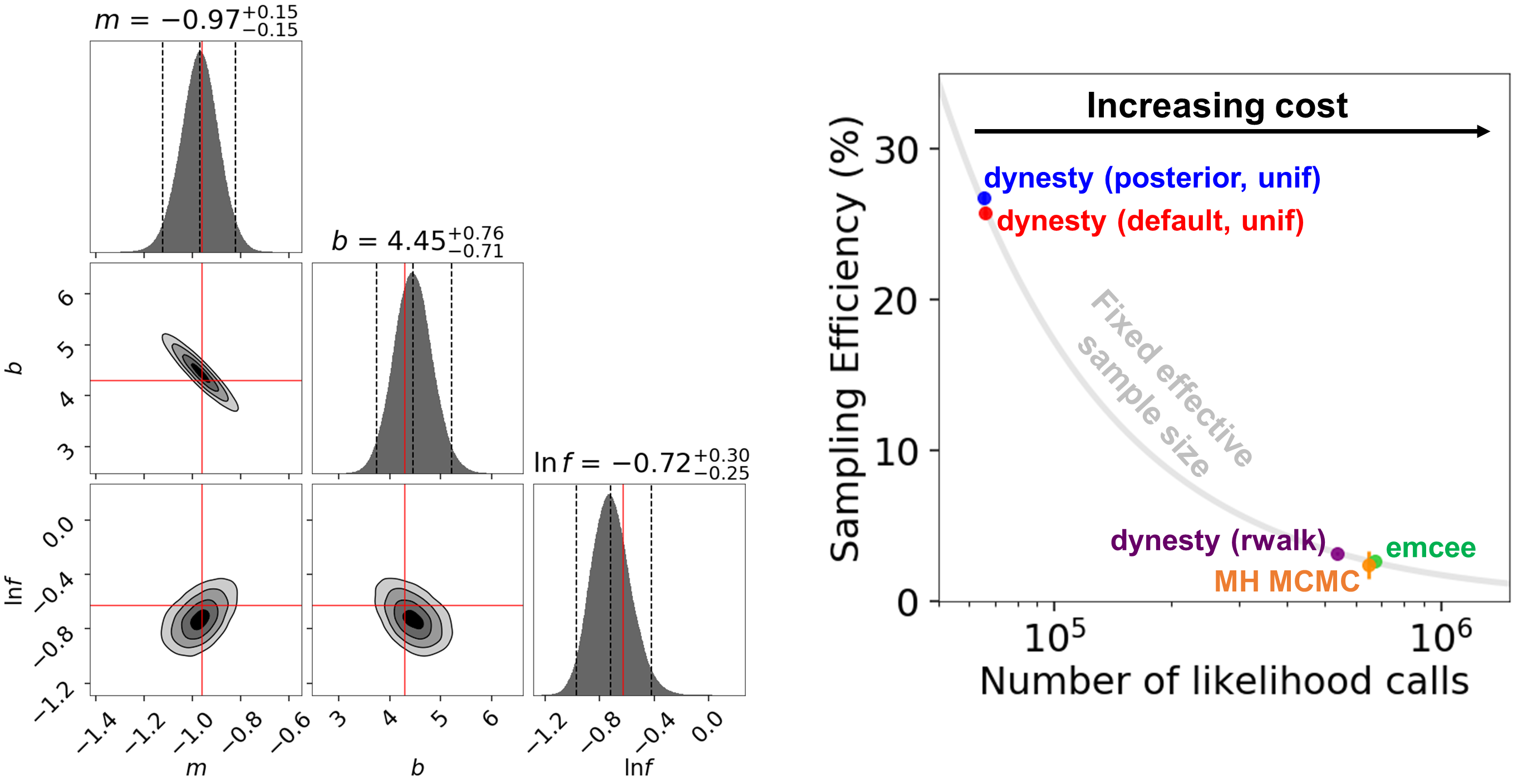}
    \caption{Comparison between {\dynesty} and common MCMC
    alternatives inferring the slope $m$, intercept $b$, and (log-)fractional
    uncertainty $\ln f$ in a simple linear regression problem.
    See \S\ref{subsec:mcmc} for additional details. 
    \textit{Left}: A corner plot showing the 1-D and 2-D marginalized
    posteriors for the slope $m$, intercept $b$, and (log-)fractional
    uncertainty $\ln f$, with their true values in red. The 2.5\%, 50\%,
    and 97.5\% percentiles (i.e. the 2-sigma credible region) are
    shown as vertical dashed lines.
    We see the posterior is well-constrained and roughly Gaussian. 
    \textit{Right}: The posterior sampling efficiency
    (i.e. the fraction of independent posterior samples
    generated per likelihood call)
    for {\dynesty}, {\emcee}, and simple MH MCMC plotted
    as a function of the total number of likelihood function calls.
    The predicted efficiency for a fixed effective 
    sample size is shown in gray.
    We see that {\dynesty} optimized for posterior
    estimation using uniform sampling (blue) can be up to 10x more
    efficient than {\emcee} or MH MCMC at generating independent samples 
    from the posterior. As expected, decreasing the emphasis on posterior
    vs evidence estimation to 80\% (red) or using a less
    efficient but more flexible sampling method such as random walks (right)
    also reduces the overall efficiency.}
    \label{fig:mcmc}
\end{figure*}

\subsection{Exponential Wave}
\label{subsec:expwave}

We next apply {\dynesty} to a signal reconstruction
problem with multiple modes and periodic boundary conditions.
Our model is a transformed periodic single 
from $0$ to $2\pi$:
\begin{equation}
    y(x) 
    = \exp\left[ n_a \sin(f_a x + p_a) 
    + n_b \sin(f_b x + p_b) \right]
\end{equation}
where we observe noisy data points drawn from
\begin{equation}
    \hat{y}(x) \sim \Normal{y(x)}{\sigma^2}
\end{equation}
The likelihood for this model is Gaussian
over the corresponding observed datapoints such that
\begin{equation}
    \ln \likelihood(\params) 
    = -\frac{1}{2} \sum_{i=1}^{N} \ln\left(2\pi\sigma^2\right) +
    \frac{\left[\hat{y}_i - y(x_i|\params)\right]^2}{\sigma^2}
\end{equation}
and has
seven free parameters: two controlling the relevant
amplitudes ($n_a$, $n_b$), two controlling the
frequencies ($f_a$, $f_b$), two controlling the phases
($p_a$, $p_b$), and one controlling the scatter $\sigma$.

We take our true model parameters to be $f_a=1.05$, $f_b=4.2$,
$n_a=0.8$, $n_b=0.3$, $p_a=0.1$, $p_b=2.4$, and $\sigma=0.2$
so that a solution is close to the boundary. We assign our
prior to be uniform or log-uniform in all parameters with
$\log n_a \in [-2, 2)$, $\log f_a \in [-2, 2)$, $p_a \in [0, 2\pi)$, 
$\log n_b \in [-2, 2)$, $\log f_b \in [-2, 2)$, $p_b \in [0, 2\pi)$, 
and $\log \sigma \in [-2, 0)$, where the priors
in $p_a$ and $p_b$ are periodic.

We illustrate {\dynesty}'s performance on this problem
in Figure \ref{fig:expwave}. We find {\dynesty} is able
to robustly recover both modes in this problem,
including the solution near the boundary.

\subsection{200-D Gaussian}
\label{subsec:200d}

We next examine {\dynesty}'s behavior in higher dimensions
by testing its performance on a 200-D multivariate Gaussian
likelihood with mean $\boldsymbol{\mu} = \mathbf{0}$ and
covariance $\cov = \mathbf{I}$ where $\mathbf{I}$ is the identity
matrix. We assign an identical prior (iid Gaussian with
$\boldsymbol{\mu} = \mathbf{0}$ and $\cov = \mathbf{I}$), such
that the posterior will also be iid Gaussian with mean
$\boldsymbol{\mu} = \mathbf{0}$ but with covariance
$\cov = (1/2) \, \mathbf{I}$. 

We sample from this distribution using Hamiltonian
Slice Sampling with the analytic log-likelihood gradient.
To further highlight the efficiency of these proposals to
explore the posterior, we use a small ($K=50$) number of live points
so that we are highly undersampled relative to the 200-D space.
Since {\dynesty} by default uses the empirical covariance
(i.e. the MLE estimate) to construct any bounding ellipsoids,
this process is dominated by shot noise
that can substantially affect the covariance. 
We consequently impose no bounding distribution (which happens
to also be optimal for this problem).

As shown in Figure \ref{fig:200d},
we find {\dynesty} is able to achieve unbiased recovery of the
mean, covariance, and evidence under these conditions.
The typical sampling efficiency we achieve for this problem
is roughly $0.1\%$ (i.e. 1000 likelihood calls per iteration), 
which translates to roughly 5 per dimension.

\begin{figure*}
	\includegraphics[width=0.98\textwidth]{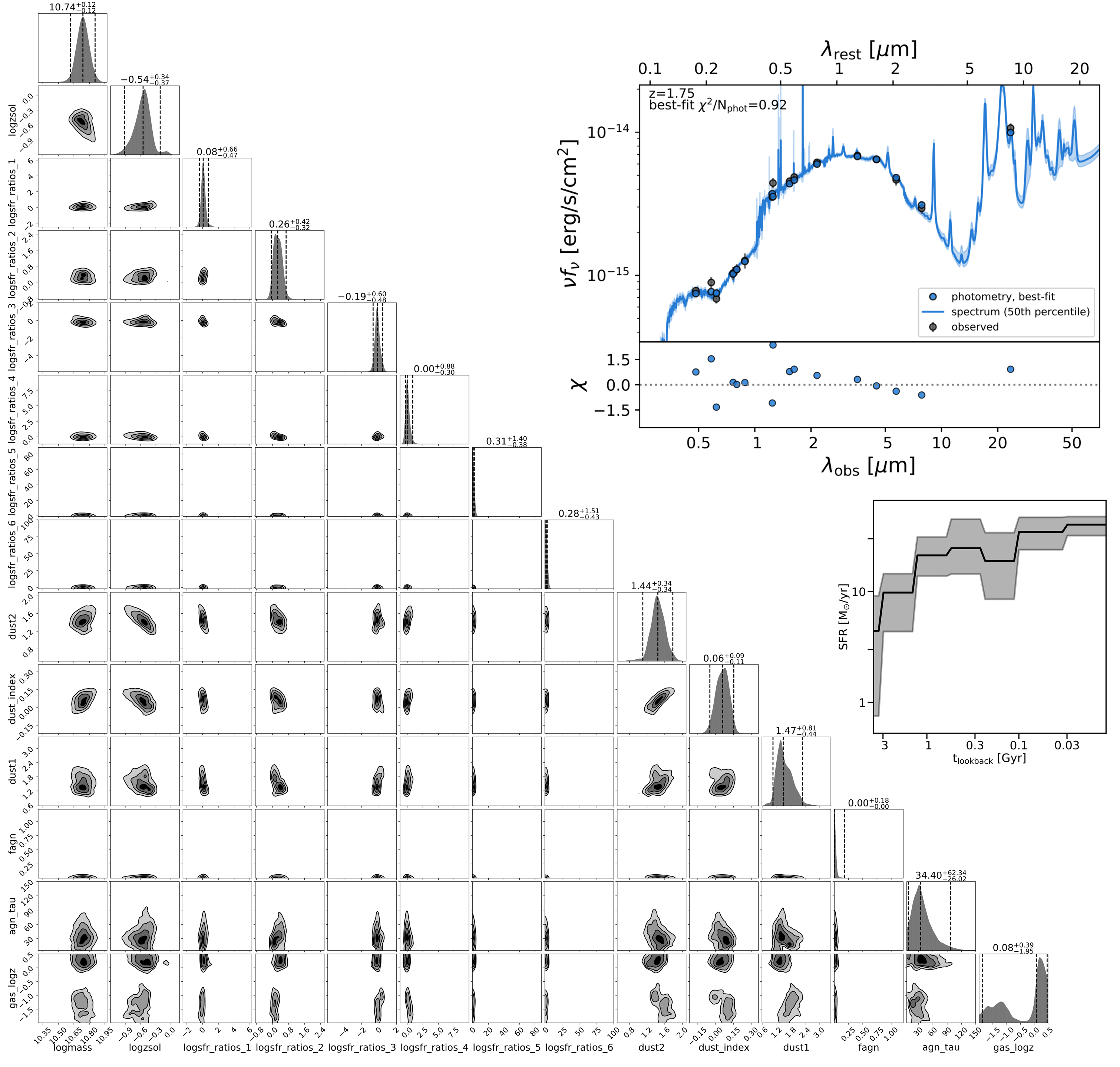}
    \caption{Galaxy SED for object AEGIS 17 from the 3D-HST
    survey modeled with \texttt{Prospector} using {\dynesty}. 
    \textit{Left}: A corner plot showing the 1-D and 2-D marginalized
    posteriors for the 14-parameter galaxy model. The 2.5\%, 50\%,
    and 97.5\% percentiles (i.e. the 2-sigma credible region) are
    shown as vertical dashed lines. The posterior
    includes a bi-modal solution for the gas-phase metallicity. 
    \textit{Top right}: The modeled galaxy SED marginalized over 
    the posterior. The 1-sigma (16-84\% credible region) is also
    shown, along with the error-normalized residuals. The
    underlying model provides a reasonable fit to the observed data.  
    \textit{Right middle}: The median reconstructed star formation history
    as a function of look-back time along with the associated 
    16-84\% credible region.}
    \label{fig:prospector}
\end{figure*}

\begin{figure*}
	\includegraphics[width=\textwidth]{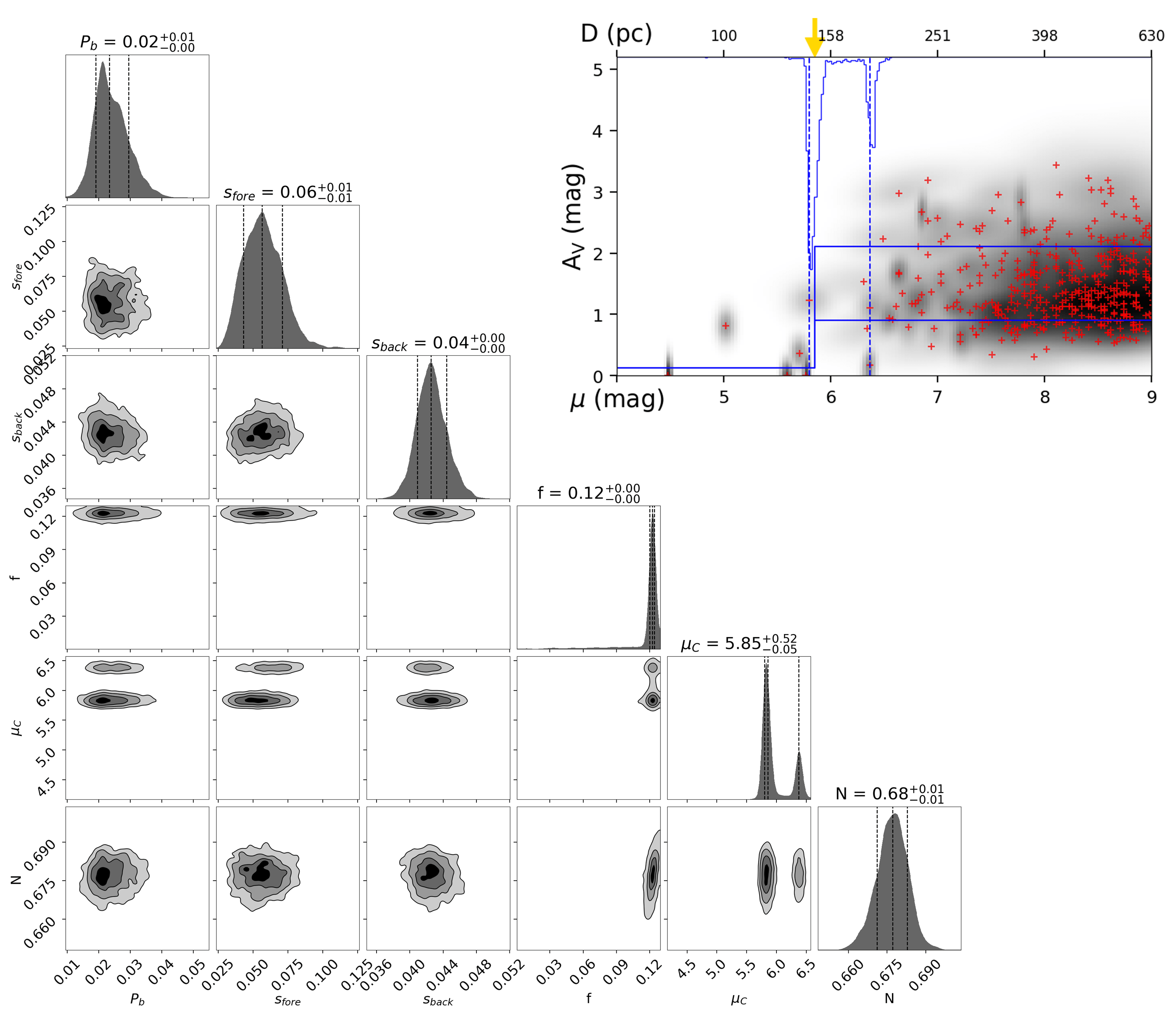}
    \caption{Line-of-sight dust extinction (reddening) model
    for a sight-line in the Chameleon molecular cloud estimated
    with {\dynesty}. 
    \textit{Left}: A corner plot showing the 1-D and 2-D marginalized
    posteriors for the 6-parameter line-of-sight model. The 16\%, 50\%,
    and 84\% percentiles (i.e. the 1-sigma credible region) are
    shown as vertical dashed lines. The posterior
    includes a bi-modal solution for the cloud distance $\mu_C$
    as well as an extended tail for the foreground dust reddening
    $f$. 
    \textit{Top right}: The line-of-sight model from the
    estimated posterior. Individual distance-extinction 
    posteriors for stars used in the fit as shown in grayscale, with
    most probable distance and extinction shown as a red cross.
    The blue line shows the typical extinction profile inferred for the sightline. 
    The range of distance estimates is shown as the inverted blue 
    histogram at the top of each panel, with the median cloud
    distance marked via the vertical blue line and yellow arrow and the
    16-84\% credible ranges marked via the vertical blue dashed lines.
    The horizontal blue lines show the estimated 1-sigma scatter
    in extinction behind the cloud.}
    \label{fig:dust}
\end{figure*}

\subsection{Comparison to MCMC}
\label{subsec:mcmc}

Nested Sampling and MCMC sampling are different
tools designed for different types of problems. 
Here we perform a limited comparison to highlight the 
advantages/disadvantages of each methodology.

We consider a simple linear regression problem where
our model is
\begin{equation}
    y(x) = mx + b
\end{equation}
and we observe noisy data from
\begin{equation}
    \hat{y}_i \sim \Normal{y(x_i)}{\sigma^2_i + [fy(x_i)]^2}
\end{equation}
where $\sigma^2_i$ is the measured variance and $f$ corresponds
to an additional fractional systematic uncertainty that we would
like to infer in addition to $m$ and $b$. The likelihood
is again Gaussian:
\begin{align}
    \ln \likelihood(m,b,f) 
    = -\frac{1}{2} \sum_{i=1}^{N} 
    &\ln\left[2\pi(\sigma^2_i + f^2(mx_i+b)^2)\right] \nonumber \\
    &+ \frac{\left[\hat{y}_i - (mx_i+b)\right]^2}{\sigma^2 + f^2(mx_i+b)^2}
\end{align}

This problem
is unimodal and only has three parameters, making it very
tractable to both Nested Sampling and MCMC methods.

We choose our priors to be uniform so that
$m \in [-5, 0.5)$, $b \in [0, 10)$, and $\ln f \in [-10, 1]$,
which are substantially broader than the likelihood distribution
but not so broad that the runtime of {\dynesty} will be dominated
merely integrating over the prior. 

We run {\dynesty} in three configurations to sample from this posterior
distribution, using the default settings whenever possible to highlight
performance in a ``typical'' use case.
First, we set the weight function to give the posterior 
100\% of the importance when allocating live points
in order to imitate MCMC-like behavior. Then, we revert to
the default 80\%/20\% posterior/evidence weighting scheme
to see how much our posterior estimate degrades as we spend 
a larger fraction of runtime trying to improve our evidence estimates. 
Finally, we switch out the default sampling mode (uniform sampling)
for random walks to forcibly decrease the overall sampling efficiency.

We compare these results to two MCMC alternatives. The
first is {\emcee} \citep{foremanmackey+13},
which is a common MCMC sampler
used in astronomical analyses today. We opt to run it in its default
configuration, which uses the ``stretch move''
from \citep{goodmanweare10} to make proposals, 
with $K=50$ walkers. We initialize the walkers
around the \textit{maximum-a-posteriori} (MAP) solution based on the
estimated covariance. We remove the first 300 samples from
the chain to account for burn-in but do not count these
``wasted'' samples when computing the overall sampling efficiency.

The second alternative is a standard MH MCMC sampler with
a Gaussian proposal distribution. We take the covariance to
be the same as that of the posterior distribution determined from
the final set of weighted {\dynesty} samples to create a
relatively optimal proposal distribution.
We then run with an identical setup to {\emcee}
(i.e. $K=50$ chains initialized around MAP solution) to maintain consistency
between approaches.

The metric we use to compare between methods is the overall
``sampling efficiency'', which we define to be 
the ratio of the estimated effective sample size (ESS) $N_{\rm ESS}$
relative to the number of likelihood calls $N_{\rm call}$:
\begin{equation}
    f_{\rm samp} \equiv \frac{N_{\rm ESS}}{N_{\rm call}}
\end{equation}
For {\dynesty}, since the samples are all independent but
assigned varying importance weights,
we choose to estimate the ESS by counting the number
of \textit{unique} samples after using systematic resampling
to redraw a set up equally-weighted samples.\footnote{
Using multinomial resampling, which introduces additional
sampling noise \citep{douc+05,hol+06}, reduces the relative ESS by roughly 25\%
but does not affect our overall conclusions.}

For the MCMC approaches, we use the standard definition of ESS as
\begin{equation}
    N_{\rm ESS} = \frac{N}{\tau}
\end{equation}
where $\tau$ is the auto-correlation averaged over all the chains.
Since $\tau$ is computed for each parameter, to be conservative 
we set the value used to compute the ESS to be the maximum value.
These choices tend to decrease the ESS by $\sim 25\%$ relative
to more optimistic ones but does not affect our overall conclusions.

We compare the five different cases above and
summarize the results from 25 independent trials
in Figure \ref{fig:mcmc}. In all cases,
we try to generate enough samples to give similar ESS between
each approach based on {\dynesty}'s default stopping criterion,
which gives $N_{\rm ESS} \sim 17000$.
We see that {\dynesty} with uniform sampling within multiple
bounding ellipsoids is roughly an order of magnitude
more efficient at generating independent samples in this problem
than MH MCMC and {\emcee}. {\dynesty} using random walks
(i.e. running MH MCMC internally) gives efficiencies that are much more
comparable to the two MCMC implementations.

As discussed earlier, all methods experience some amount of
overhead transitioning from the prior-dominated to
posterior-dominated region. While this leads to
$\lesssim 5\%$ of samples being discarded for burn-in 
for the MCMC cases, it leads to a reduction in the ESS
of $\sim 25\%$ for {\dynesty}. The fact that {\dynesty}
performs well even in this case illustrates how important
Dynamic Nested Sampling is for ensuring samples
are efficiently allocated during runtime.

This result highlights
the basic argument first outlined in \S\ref{sec:nested},
illustrating that using Nested Sampling to
sample from many simpler distributions
in turn can sometimes be more effective than trying to sample
from the posterior distribution directly with MCMC.
In general, Nested Sampling performs well in cases
like these where the likelihood varies smoothly in a given region
and the prior has reasonable bounds.
In other cases where the prior is large or fewer samples
from the posterior are needed, MCMC methods are more than sufficient.

\section{Applications}
\label{sec:applied}

In addition to the toy problems in \S\ref{sec:tests}, {\dynesty}
has also been applied in several packages and ongoing studies
and shown to perform well when applied to real astronomical
analyses. These include applications analyzing gravitational
waves \citep{ashton+18}, 
exoplanets \citep{diamondlowe+18,espinoza+18,gunther+19},
transients \citep{guillochon+18},
galaxies \citep{leja+18b,leja+18c}, 
and 3-D dust mapping \citep{zucker+18,zucker+19}. 
We highlight two of these applications below that the
author has been personally involved in.

In \citet{leja+18c}, the authors modeled roughly 60k galaxy spectral energy
distributions (SEDs) from the 3D-HST survey \citep{brammer+12} over a redshift
range of $0.5 < z < 2.5$. To conduct this analysis, they used the
Bayesian SED fitting code \texttt{Prospector} (Johnson et al. in prep.),
utilizing {\dynesty} as their primary sampler, to sample from a 14-parameter model
involving stellar mass, a non-parametric star formation history, stellar and
gas metallicites, dust properties, and contributions from 
possible Active Galactic Nuclei. Compared to previous studies where
{\emcee} had been used to sample from the posterior
\citep{leja+17,leja+18a}, the authors found that {\dynesty} provided
over an order of magnitude more efficient sampling and was able to
characterize a wide variety of posteriors. The results
for a typical galaxy are shown in Figure \ref{fig:prospector}.

In \citet{zucker+19}, the authors used a combination of distance and reddening
estimates to nearby stars from SED modeling (Speagle et al. in prep.)
and Gaia parallax measurements \citep{gaia+18} to derive distances to dozens
of local molecular clouds. The distances to these clouds, however,
are sensitive to the number and distribution of
stars immediately in front of them as these stars help
constrain the location of the ``jump'' in dust extinction associated with the
cloud. In cases where there are only a small number of foreground stars, this
constraint can be quite weak, leading to extended
posteriors with multi-modal solutions. This, along with the
overall performance illustrated in Figure \ref{fig:mcmc},
motivated the use of {\dynesty} to sample from the 6-parameter
cloud distance model used in the analysis. We highlight one such
multi-modal case in Chameleon in Figure \ref{fig:dust}.

These examples, along with others listed earlier,
are large-scale professional applications of {\dynesty} that
illustrate {\dynesty} can work well in theory and in practice.

\section{Conclusion}
\label{sec:conclusion}

With Bayesian inference techniques now a large part of modern
astronomical analyses, it has become increasingly
important to develop and provide tools to the
community that can help to ``bridge the gap''
between writing the underlying model and estimating the
corresponding posterior $\posterior(\params)$. Tools such
as {\emcee}, {\multinest}, and {\polychord}, which provide Markov Chain
Monte Carlo and Nested Sampling implementations,
have been heavily used and highly cited.

In this paper we presented an overview of {\dynesty},
a public, open-source, Python package that implements 
Dynamic Nested Sampling to enable
flexible Bayesian inference over complex, multi-modal distributions.
Building on previous work in the literature,
we described the basics behind the Dyamic Nested Sampling 
approaches employed in the code, how we implement them,
and how we use a variety of bounding and sampling methods
to enable efficient inference. We then showcased
{\dynesty}'s performance on several toy
problems as well as real astronomical application, highlighting
its ability to estimate challenging posterior distributions
both in theory and in practice.

While we have shown {\dynesty} can perform
similarly or better than existing MCMC approaches in one simple case,
the real test for any package is based on users applying
it to their analysis problems. We hope that {\dynesty} will
prove useful to the community and help facilitate 
exciting new science over the coming years.

\section*{Acknowledgements}

JSS is grateful to Rebecca Bleich for her support and patience.

This project is the culmination of many individual efforts,
not all of whom can be thanked here. 
First and foremost, JSS would like to thank
Daniel Eisenstein, Charlie Conroy, and Doug Finkbeiner
for their patience while he pursued this project, 
Catherine Zucker for her constant stream of feedback during development,
and Johannes Buchner for incredibly insightful and inspiring conversations.
JSS would also like to thank Johannes Buchner, Hannah Diamond-Lowe,
Daniel Eisenstein, Daniel Foreman-Mackey, Will Handley, Ben Johnson,
Joel Leja, Locke Patton, and Catherine Zucker for feedback on early drafts 
that substantially improved the quality of this work.
JSS would further like to thank Johannes Buchner, Phil Cargile,
Ben Cook, James Guillochon, and Ben Johnson
for their direct and indirect contributions to the {\dynesty} codebase,
as well as Kyle Barbary and collaborators for their contributions to {\nestle}
(upon which {\dynesty} was initially based).
JSS is also grateful to many beta-testers who provided invaluable feedback
during {\dynesty}'s development and suffered through many bugfixes,
including (but not limited to)
Gregory Ashton, Ana Bonaca, Phil Cargile, Tansu Daylan,
Hannah Diamond-Lowe, Philipp Eller, Jonathan Fraine, Maximilian G\"unther,
Daniela Huppenkothen, Joel Leja, Sandro Tacchella, Ashley Villar, 
Catherine Zucker, and Joe Zuntz.

This work has benefited from several software packages including
\texttt{numpy} \citep{vanderwalk+11}, \texttt{scipy} \citep{oliphant07}, 
\texttt{matplotlib} \citep{hunter07}, and \texttt{corner} \citep{foremanmackey16}.




\bibliographystyle{mnras}
\bibliography{ref}



\appendix

\section{Detailed Nested Sampling Results}
\label{ap:nested}

While we presented a broad overview of Nested Sampling
in the main text, we glossed over much of the statistical
background. We include more detailed results and discussion below.

The outline of these results are as follows.
In \S\ref{subap:nested_setup} we outline the basic setup for
Nested Sampling. 
In \S\ref{subap:nested_single} we derive 
statistical properties in the single live point case.
In \S\ref{subap:nested_parallel} we discuss the process
of utilizing multiple live points. 
In \S\ref{subap:nested_many}
we derive properties in the many live point case.
In \S\ref{subap:nested_dynamic} we extend these
results to encompass varying numbers of live points.
Finally, in \S\ref{subap:nested_errors} we discuss various error
properties of Nested Sampling as well as schemes
to estimate them.

\subsection{Setup}
\label{subap:nested_setup}

Following \citet{skilling06}, \cite{feroz+13}, and others,
we start by (re-)defining Bayes Rule
\begin{equation}
    \posterior(\params)
    = \frac{\likelihood(\params) \prior(\params)}{\evidence}
\end{equation}
where $\posterior(\params)$ is the posterior, $\likelihood(\params)$ is the
likelihood, $\prior(\params)$ is the prior, and
\begin{equation}
    \evidence_M
    = \int_{\Omega_{\params}} \likelihood(\params) \prior(\params)
    \deriv \params
\end{equation}
is the evidence.

To evaluate this integral, Nested Sampling seeks to transform it from one over
position $\params$ to one over prior volume $X$ where
\begin{equation}
    X(\lambda) 
    \equiv \int_{\Omega_{\params} \,{\rm s.t.} \likelihood(\params) > \lambda}
    \prior(\params) \deriv \params
    \equiv \int_{\Omega_{\params}} \prior_\lambda(\params) \deriv \params
\end{equation}
defines the prior volume within a given 
iso-likelihood contour of level $\lambda$,
assuming our priors are integrable, and
\begin{equation}
    \prior_\lambda(\params) 
    \equiv
    \begin{cases}
    \prior(\params) / X(\lambda) & \likelihood(\params) \geq \lambda \\
    0 & \likelihood(\params) < \lambda
    \end{cases}
\end{equation}
is the constrained prior. Note that $X \in (0, 1]$ since the integral
over the entire prior is $x(\lambda=0)=1$ while the value as 
$\lambda \rightarrow \infty$ should approach $0$ if the maximum-likelihood value
$\likelihood_{\max}$ is a singular point.

Since $\lambda \in [0, \infty)$, this allows us to 
redefine the evidence integral as
\begin{equation}
    \evidence = \int_0^\infty X(\lambda) \deriv \lambda
\end{equation}
Provided the inverse $\likelihood(X)$ of $X(\likelihood(\params)=\lambda)$ 
exists (i.e. there are no flat ``slabs'' of likelihood anywhere, 
only contours), we can rewrite this integral in terms 
of the prior volume associated with
a particular iso-likelihood contour:
\begin{equation}
    \evidence = \int_0^1 \likelihood(X) \deriv X
\end{equation}
This is now a 1-D integral over $X$ that we can approximate using
a discrete set of $N$ points using, e.g., a Riemann sum
\begin{equation}
    \hat{\evidence} 
    = \sum_{i=1}^{N} \likelihood(\params_i) \times (X_i - X_{i-1})
    \equiv \sum_{i=1}^{N} p(\params_i)
\end{equation}
where $X_0 = 1$ and $p(\params_i)$ is the (un-normalized) importance weight.
These values can also be used to approximate the posterior:
\begin{equation}
    \hat{\posterior}(\params) 
    = \frac{\sum_{i=1}^{N} p(\params_i) \delta(\params_i)}
    {\sum_{i=1}^{N} p(\params_i)}
\end{equation}

\subsection{Using a Single Live Point}
\label{subap:nested_single}

Unfortunately, the exact value of $X(\lambda)$ at a given likelihood
level $\lambda = \likelihood(\params)$ is unknown. We can, however,
construct an estimator $\hat{X}$ with a
known statistical distribution. Looking
back at the definition of the prior volume $X(\likelihood)$, 
we see that it defines a cumulative distribution function (CDF)
over $\likelihood$. We can then define the associated probability density
function (PDF) for $\likelihood$ as
\begin{equation}
    P(\likelihood)
    \equiv \frac{\deriv X(\likelihood)}{\deriv \likelihood} 
    = \frac{\deriv}{\deriv \likelihood} 
    \int_{\Omega_{\params}} \prior_\likelihood(\params) \deriv \params
\end{equation}

Assuming we can sample $\likelihood$ from its 
PDF $P(\likelihood)$,
we can use the Probability Integral Transform (PIT)
to subsequently constrain
the distribution of $X(\likelihood)$. In other words:
\begin{equation}
    \likelihood' \sim P(\likelihood) 
    \quad\Rightarrow\quad X(\likelihood') \sim \Unif
\end{equation}
where $X \sim f(X)$ notation implies the random variable $X$ is drawn from $f(X)$
and $\Unif$ is the standard Uniform distribution (i.e. flat from $0$ to $1$).
This can be directly extended to cases where we are interested in sampling
relative to a given threshold $\lambda$ as
\begin{equation}
    \likelihood' \sim P(\likelihood | \likelihood > \lambda) 
    \quad\Rightarrow\quad \frac{X(\likelihood')}{X(\lambda)} \sim \Unif
\end{equation}

While this does not appear to make things any easier, it actually helps
us out enormously. That's because, at fixed $\lambda$,
$X(\lambda)$ is actually a CDF over the constrained prior
$\prior_\lambda(\params)$. That means we can bypass $\lambda$
and $P(\likelihood)$ altogether and 
just sample from $\prior_\lambda(\params)$ directly
to satisfy the PIT:
\begin{equation}
    \params' \sim \prior_\lambda(\params) 
    \quad\Rightarrow\quad 
    \frac{X(\likelihood(\params'))}{X(\lambda)} \sim \Unif
\end{equation}
Various methods for sampling from the 
constrained prior $\prior_\lambda(\params)$
subject to a suitable prior transform $\ptform$ 
(see \S\ref{subsec:nested_samples})
are outlined in \S\ref{sec:methods}.

Before moving on, we want to quickly note that while the
above scheme is \textit{sufficient} for generating
values of $\likelihood' \sim P(\likelihood)$ it is by no means
\textit{necessary}. As a counter-example, we can imagine
a function $f(t)\rightarrow\params_t$ 
that traces out a singular path through
the distribution with support over 
$\likelihood(f) \in [\likelihood^{\min}, \likelihood^{\max}]$.
Let us furthermore assume that we construct $f(t)$ such that
we spend more ``time'' $t$ where the likelihood PDF is higher
so that the PDF $P(t) \propto P(\likelihood(\params_t))$.
Finally, let's define the constrained function $f_\lambda(t)$
to simply be the portion of the path with 
$\likelihood(\params_t)>\lambda$.
While this path by no means encompasses the prior, it is clear that
\begin{equation}
    t' \sim f_\lambda(t) 
    \quad\Rightarrow\quad 
    \frac{X(\likelihood(\params_{t'}))}{X(\lambda)} \sim \Unif
\end{equation}
This result proves we can in theory satisfy the PIT for Nested Sampling
using correlated samples provided they
probe enough of the \textit{local} portion of the prior to obtain
sufficient coverage over the range of possible likelihoods
\citep[see also][]{salomone+18}. It also provides support
for why Nested Sampling works so well in practice even when
samples are not fully independent.

For a given prior volume $X_{i-1}$ associated with
a given likelihood level $\lambda_{i-1} = \likelihood(\params_{i-1})$
after $i-1$ iterations of this procedure, this implies the
current prior volume $X_i$ will be
\begin{equation}
    \hat{X}_i = U_i \hat{X}_{i-1} = \prod_{j=1}^{i} U_j
\end{equation}
where
\begin{equation}
    U_1, \dots, U_i \stackrel{\iid}{\sim} \Unif
\end{equation}
are independent and identically distributed (iid) random variables
drawn from the standard Uniform distribution and we have taken 
$X_0 \equiv X(\lambda=0) = 1$.
As we do not actually know the values of $U_1,\dots,U_j$, we consider
$\hat{X}_i$ to be a noisy estimator of $X_i$. 

While sampling, we obviously need to assign a value for $\hat{X}_i$
to determine, e.g., whether to stop. While we can easily simulate
random values of $U_1,\dots,U_i$, if we want these values to be
consistent then a reasonable choice
is the expectation value (arithmetic mean):
\begin{equation}
    \mean{\hat{X}_i} 
    = \mean{\prod_{j=1}^{i} U_j} 
    = \prod_{j=1}^{i} \mean{U_j}
    = \left(\frac{1}{2}\right)^{i}
\end{equation}
Alternately, we might also be interested in the expectation
value of $\ln \hat{X}_i$ (geometric mean):
\begin{equation}
    \mean{\ln \hat{X}_i} 
    = \sum_{j=1}^{i} \mean{\ln U_j}
    \sim -\sum_{j=1}^{i} \mean{E_j}
    = -i
\end{equation}
where we have used the fact that
\begin{equation}
    U \sim \Unif
    \Rightarrow -\ln U \sim \Expo
\end{equation}
where $\Expo$ is the standard Exponential
distribution and
\begin{equation}
    E_1,\dots,E_i \stackrel{\iid}{\sim} \Expo
\end{equation}

Various stopping criteria are discussed in
the main text (\S\ref{subsec:nested_stop}
and \S\ref{subsec:dynamic_stop})
and so are not discussed further here.

\subsection{Combining Live Points}
\label{subap:nested_parallel}

Following \citet{higson+17a}, 
let's consider the case where we have two independent
live points following the basic sampling approach described
above. These each form a set of samples with increasing likelihood
\begin{align*}
    \likelihood_{N_1}^{[1]} > \dots > \likelihood_1^{[1]} > 0 \\
    \likelihood_{N_2}^{[2]} > \dots > \likelihood_1^{[2]} > 0
\end{align*}
where the $[\cdot]$ superscript notation indicates the 
index of the associated live point. 
We now want to ``merge'' these two sets of ordered
samples together to get a single hypothetical ordered list:
\begin{align*}
    &\likelihood_{N_1}^{[1]} > \likelihood_{N_2}^{[2]} > \dots >
    \likelihood_{2}^{[1]} > \likelihood_{2}^{[2]} >
    \likelihood_{1}^{[2]} > \likelihood_{1}^{[1]} > 0 \\
    &\rightarrow \likelihood_N > \dots > \likelihood_1 > 0
\end{align*}
where $N = N_1 + N_2$.

Independently, we know that the prior volume at a given iteration 
for each live point is just
\begin{equation*}
    \hat{X}_i^{[j]} = \prod_{n=1}^{i} U_n^{[j]}
\end{equation*}
What we want to know, however, is the distribution of $\hat{X}_i$
of the \textit{merged} list. 
Considering each sample independently implies $X_2 = X_1^{[2]}$
and $X_1 = X_1^{[1]}$ follow the same distribution (i.e. the first sampled
prior volume for each run is similarly distributed).
However, considering them
\textit{together} (based on the merged list) implies $X_2 = X_1^{[2]}$ is
\textit{strictly less than} $X_1 = X_1^{[1]}$ 
since $\likelihood_2 > \likelihood_1$.
This tells us that $\hat{X}_i$ \textit{cannot} follow the same distribution
from the associated independent runs that comprise it.

With this finding in hand, we now consider
an approach for sampling from the prior volume using two live points. 
At each iteration $i$, we remove the one with the lowest likelihood
$\lambda = \likelihood^{\min}_i$ and replace it with a new point sampled from the
constrained prior $\prior_\lambda(\params)$. After $N$ iterations, we will
end up with a sorted list of likelihoods
$\likelihood_N > \dots > \likelihood_1 > 0$. If, however, we look at each live point
individually (i.e. ignoring the other live point), we would find that each live point's
evolution would comprise a list of independent samples with ordered likelihoods
that would each be identical to
$\likelihood_{N_1}^{[1]} > \dots > \likelihood_1^{[1]} > 0$ and
$\likelihood_{N_2}^{[2]} > \dots > \likelihood_1^{[2]} > 0$, respectively!
Therefore, we see that this procedure for sampling with two live points
is identical to combining two sets of independent samples
derived using one live point each.

The above procedure can be immediately generalized to $K$ live points,
producing the (Static) Nested Sampling 
procedure outlined in Algorithm \ref{alg:static}. We will return
to this duality between $K$ independent Nested Sampling runs and 
a single Nested Sampling run with $K$ live points 
in \S\ref{subap:nested_errors}.

\subsection{Using Many Live Points}
\label{subap:nested_many}

Now that we have established a procedure for running Nested
Sampling with $K$ live points, we need to characterize how this
affects our estimates $\hat{X}_i$ of the prior volume. At any
given iteration $i$, we know that the current 
set of prior volumes $\lbrace X_i^{[1]}, \dots, X_i^{[K]} \rbrace$
associated with our $K$ live points are uniformly distributed
within the prior volume from the previous iteration $X_{i-1}$
so that
\begin{equation}
    X_i^{[j]} = U^{[j]} X_{i-1}
\end{equation}
where
\begin{equation}
    U^{[1]},\dots,U^{[K]} \stackrel{\iid}{\sim} \Unif
\end{equation}

We are now want to replace the live point with the lowest likelihood
$\likelihood_i^{\min}$ corresponding to the largest prior volume.
This means we are now interested in the \textit{ordered} list of
prior volumes
\begin{equation}
    X_i^{(j)} = U^{(j)} X_{i-1}
\end{equation}
where $(j)$ now indicates the position in the \textit{ordered list}
(from smallest to largest) rather than the live point index $[j]$
and
\begin{equation}
    U^{(j)} 
    = \min_j\left(\left\lbrace U^{[1]},\dots,U^{[K]} \right\rbrace\right)
\end{equation}
is the $j$th standard uniform order statistic,
where $\min_j$ selects the $j$th smallest point (so $j=1$ is the smallest
and $j=K$ is the largest).

Using the Reny\'i Representation, it can be shown
that we can represent the \textit{joint} distribution
of our $K$ standard uniform order statistics 
$\lbrace U^{(1)}, \dots, U^{(K)} \rbrace$ such that \citep{nagaraja06}:
\begin{equation}
    U^{(j)}
    = \frac{\sum_{n=1}^{j} E_n}{\sum_{n=1}^{K+1} E_n}
\end{equation}
where
\begin{equation}
    E_1,\dots,E_{K+1} \stackrel{\iid}{\sim} \Expo
\end{equation}
The marginal distribution for $U^{(j)}$ is then \citep{blitzsteinhwang14}:
\begin{equation}
    U^{(j)} \sim \Beta{j}{K+1-j}
\end{equation}
where $\Beta{\alpha}{\beta}$ is the Beta distribution.

Using these results, we see that the prior volume
based on $K$ live points at iteration $i$ evolves as
\begin{equation}
    \hat{X}_i = \prod_{j=1}^{i} U_j^{(K)}
\end{equation}
where $U_1^{(K)},\dots,U_i^{(K)}$ are iid draws of the
$K$th standard uniform order statistic with marginal
distribution $\Beta{K}{1}$.
The arithmetic mean is
\begin{equation}
    \mean{\hat{X}_i} 
    = \prod_{j=1}^{i} \mean{U_j^{(K)}}
    = \left(\frac{K}{K+1}\right)^{i}
\end{equation}
The geometric mean is
\begin{equation}
    \mean{\ln \hat{X}_i} 
    = \sum_{j=1}^{i} \mean{\ln U_j^{(K)}}
    = -\frac{i}{K}
\end{equation}

As discussed in \S\ref{subsec:nested_volume}, after we terminate
sampling we can add the final set of $K$ live points to our
set of $N$ samples. These will then just follow the final
set of $\lbrace U^{(1)}, \dots, U^{(K)} \rbrace$ standard
uniform order statistics relative to $\hat{X}_N$
with an arithmetic mean
\begin{equation}
    \mean{\hat{X}_{N+k}} 
    = \left(\frac{K+1-k}{K+1}\right) \left(\frac{K}{K+1}\right)^{N}
\end{equation}
and geometric mean
\begin{equation}
    \mean{\ln \hat{X}_{N+k}} 
    = -\frac{N}{K} - \left[\psi(K+1) - \psi(K+1-k)\right]
\end{equation}
where $\psi(\cdot)$ is the digamma function.

\subsection{Using a Varying Number of Live Points}
\label{subap:nested_dynamic}

As discussed in \S\ref{sec:dynamic}, there's no inherent
reason why the number of number of live points
must remain constant from iteration to iteration. Indeed,
we can interpret adding the final set of live points
to the list of samples from \S\ref{subap:nested_many}
as simply allowing the nested sampling run to continue
while continually decreasing the number of live points. 
From this viewpoint, we have $K_1=\dots=K_N=K$ live points
over iteration $i=1$ to $N$, but only $K_{N+k}=K+1-k$ live points
at iteration $i=N+k$.

The change in the number of live points also changes
the overall behavior of the Nested Sampling run before
and after adding the final set of live points.
We can highlight these by rewriting the results from
\S\ref{subap:nested_many} as:
\begin{equation}
    \ln \mean{\hat{X}_{N+k}} 
    = \underset{{\rm Exponential\:Shrinkage}}
    {\underbrace{\sum_{i=1}^{N} \ln\left(\frac{K}{K+1}\right)}}
    + \underset{{\rm Uniform\:Shrinkage}}
    {\underbrace{\sum_{j=1}^{k} \ln \left(\frac{K+1-k}{K+2-k}\right)}}
\end{equation}
This neatly decomposes the two ``modes'' in which
Nested Sampling can
traverse the prior. While ``replacing'' the worst live point
(i.e. $K_{i}=K_{i-1}$), the prior volume shrinks \textit{exponentially}
by a constant factor at each iteration.
However, when ``removing'' live points (i.e. $K_{i} < K_{i-1}$),
we instead shrink \textit{uniformly} 
by a variable factor at each iteration.

We can now generalize this behavior to the case where
$K_i$ is allowed to vary at each iteration \citep{higson+17b}.
This now generates two distinct classes of behavior. 
When $K_i \geq K_{i-1}$, we add
$K_i - K_{i-1} \geq 0$ live points to our existing
set of live points, after which we replace the
one with the worst likelihood $\likelihood_i^{\min}$.
This then gives a distribution for the prior volume
shrinkage of $\Beta{K_i}{1}$.

If $K_i < K_{i-1}$, on the other hand, 
we instead have removed $K_{i-1} - K_i$ live points
from the previous set of live points. The expected
shrinkage is then based on the associated $K_{i}$
standard uniform order statistic $U^{(K_{i})}$
from the initial set of $K_{i-1}$ values. Although
in theory we should consider cases where the number
of live points can decrease by an arbitrary amount,
in practice when following iterative schemes such as
the one outlined in Algorithm \ref{alg:dynamic_iter}
we only need to consider the case where
$K_{i-1} - K_i = 1$.

Taken together, these two types of behavior then give
a mean estimate of:
\begin{align}
    &\ln \mean{\hat{X}_{j}} 
    = \sum_{i=1}^{n_1} \ln\left(\frac{K_i}{K_i+1}\right)
    + \sum_{i=1}^{n_2} \ln \left(\frac{K_{N_1}+1-i}{K_{N_1}+2-i}\right) 
    \nonumber \\
    &+ \sum_{i=1}^{n_3} \ln\left(\frac{K_{N_2+i}}
    {K_{N_2+i}+1}\right) + \dots
    + \sum_{i=1}^{n_{M-1}} \ln\left(\frac{K_{N_{M-2}+i} + 1 - i}
    {K_{N_{M-2}+i}+1}\right) \\
    &+ \sum_{i=1}^{n_M} \ln \left(\frac{K_{N_{M-1}}+1-i}
    {K_{N_{M-1}}+2-i}\right) \nonumber
\end{align}
where $n_m$ is the number of contiguous samples for which either
exponential or uniform shrinkage dominates, 
$N_m = \sum_{k=1}^{i} n_k$ is the total number of iterations
that have occurred up to that point, and $M$ is the
number of contiguous regions prior to iteration $j = N_M$
where one mode of shrinkage dominates. Note that for illustrative
purposes here we have assumed the final
samples are experiencing uniform shrinkage.

To summarize, varying the number of live points at
each iteration simply involves dynamically 
switching between exponential and uniform shrinkage
over the course of a Nested Sampling run. While
this adds additional bookkeeping, it remains straightforward
to estimate the prior volume $\hat{X}_i$ at any particular
iteration.

\subsection{Nested Sampling Errors}
\label{subap:nested_errors}

We now turn our attention to characterizing various error properties
of Nested Sampling, following the basic approach of
\citet{higson+17a,higson+19}. Similar to other sampling approaches,
we expect some amount of ``sampling noise''
in our evidence $\hat{\evidence}$
and posterior estimates $\hat{\posterior}(\params)$
arising from the fact that we are approximating a continuous
distribution (and smooth integral) with a discrete 
set of $N$ samples.
We expect that as the number of live points at each iteration
$K_i \rightarrow \infty$ such that 
change in prior volume $X_i - X_{i-1} \rightarrow 0$
and the total number of samples
$N \rightarrow \infty$,
these sampling errors will become negligible.

Unlike other sampling approaches such as Markov Chain Monte Carlo
(MCMC), however, Nested Sampling,
contains an \textit{additional} source of 
noise arising from our use
of noisy estimators $X \rightarrow \hat{X}_i$ of the prior volume
at a given iteration $i$ \citep{skilling06}. This ``statistical noise''
translates to a noisy estimator of the importance weight
$p(\params_i) \rightarrow \hat{p}(\params_i)$, 
which in turn gives noisy estimators for our previous evidence estimate
\begin{align}
    \hat{\evidence}
    &= \sum_{i=1}^{N} \likelihood(\params_i) \times
    (X_i - X_{i-1}) \nonumber \\
    &\approx \sum_{i=1}^{N} \likelihood(\params_i) \times
    (\hat{X}_i - \hat{X}_{i-1})
    \equiv \sum_{i=1}^{N} \hat{p}(\params_i)
\end{align}
and our previous posterior estimate
\begin{equation}
    \hat{\posterior}(\params) 
    = \frac{\sum_{i=1}^{N} p(\params_i) \delta(\params_i)}
    {\sum_{i=1}^{N} p(\params_i)}
    \approx \frac{\sum_{i=1}^{N} \hat{p}(\params_i) \delta(\params_i)}
    {\sum_{i=1}^{N} \hat{p}(\params_i)}
\end{equation}
Similar with the sampling noise, we also expect
the statistical noise to become negligible as the
number of live points at each iteration 
$K_i \rightarrow \infty$ such that our estimate
$\hat{X}_i \rightarrow X_i$ and
the total number of samples
$N \rightarrow \infty$.

We can highlight the decomposition of these two noise
sources by considering trying to evaluate
the expectation value of a target function
$f(\params)$ with respect to the posterior 
\citep{chopinrobert10,higson+17a}:
\begin{equation}
    \meanwrt{f}{\posterior}
    = \int_{\Omega_{\params}} f(\params) 
    \posterior(\params) \deriv \params
    = \frac{1}{\evidence} \int_0^1 \tilde{f}(X) \likelihood(X) \deriv X
\end{equation}
where
\begin{equation}
    \tilde{f}(X)
    = \meanwrt{f(\params)|\likelihood(\params)=\likelihood(X)}{\prior}
\end{equation}
is the expectation value of $f(\params)$ on the associated iso-likelihood
contour $\likelihood(\params)=\likelihood(X)$ with respect to the prior
$\prior(\params)$. Using the same Riemann sum approximation
as \S\ref{subap:nested_setup}, Nested Sampling would
approximate this integral as:
\begin{align}
    \meanwrt{f}{\posterior}
    &\approx \sum_{i=1}^{N} \tilde{f}(X_i) 
    \frac{\likelihood(X_i) (X_i - X_{i-1})}{\evidence}
    = \sum_{i=1}^{N} \tilde{f}(X_i) p(X_i) \\
    &\approx \sum_{i=1}^{N} f(\params_i) p(X_i) \label{eq:mc} \\
    &\approx \sum_{i=1}^{N} f(\params_i) \hat{p}(\params_i) \label{eq:vol}
\end{align}
We can see the two error types enter in cleanly through the final two
approximations. In equation \eqref{eq:mc}, we introduce sampling noise
by replacing $\tilde{f}(X)$, which is averaged over the entire
iso-likelihood contour, with the estimate $f(\params_i)$ evaluated
at a single point. Then, in equation \eqref{eq:vol}, we replace the
true importance weight $p(X_i)$ at a given prior volume
with its noisy estimate $\hat{p}(\params_i)$ based on our
noisy estimators for the prior volume $\hat{X}_i$.


\subsubsection{Statistical Uncertainties}
\label{subsubap:error_stat}

In \S\ref{subap:nested_single}, \S\ref{subap:nested_many},
and \S\ref{subap:nested_dynamic},
we derived the analytic distribution for our prior volume
estimator $\hat{X}_i$ at iteration $i$ under a variety of
assumptions. While the distribution for $\hat{\evidence}$
and $\hat{\posterior}(\params)$ is not analytic,
it is straightforward to draw from them.
First, we simulate values of the prior volumes
\begin{equation}
    \hat{X}_1', \dots, \hat{X}_N' \sim P(\hat{X}_1, \dots, \hat{X}_N)
\end{equation}
by drawing a combination of $\Beta{K_i}{1}$-distributed random
variables (when $K_i \geq K_{i-1}$)
and standard uniform order statistics (when $K_i < K_{i-1}$)
and iteratively computing each $\hat{X}_i'$ using the procedures
outlined earlier. Then, we simply compute the
corresponding evidence $\hat{\evidence}'$ and 
posterior $\hat{\posterior}'(\params)$ estimates.

While we can simulate the prior volumes and trace their impact
on $\hat{\evidence}$ and $\hat{\posterior}(\params)$ explicitly,
it is also helpful to derive a rough estimate of their impact.
Since the posterior $\posterior(\params)$ can be arbitrarily complex,
we will focus on the evidence $\evidence$ for which this analysis
is more tractable.

There have previously been two main approaches for 
deriving the uncertainty, which focus either on trying
to derive $\variance{\hat{\evidence}}$ \citep{keeton11} or
$\variance{\ln \hat{\evidence}}$ \citep{skilling06}. Here we will focus
on the latter, which gives a cleaner (if less precise)
result.

We first start with the Static Nested Sampling case using
a constant number of live points $K$. To estimate the evidence
$\evidence$, we must integrate over the unnormalized posterior
$\posterior(\params) \propto \prior(\params)\likelihood(\params)$.
This occurs after a certain number of iterations $N$ have passed
given a fixed stopping criterion.

There are two factors that contribute to the overall $N$. The
first is the rate of integration:
at any given iteration $i$, the prior volume decreases by
$\Delta \ln X \approx 1/K$. As a result, it must be the case that
$N \propto 1/K$.

The second is the total amount of prior volume that needs
to be integrated over. This roughly scales as the
Kullback-Leibler (KL) divergence (i.e. ``information gain'')
between the prior $\pi(\params)$ and posterior $\posterior$
\begin{align}
    H(\posterior||\prior) 
    \equiv H 
    &\equiv \int_{\Omega_{\params}} \posterior(\params)
    \ln \frac{\posterior(\params)}{\prior(\params)} \deriv \params \\
    &= \frac{1}{\evidence} \int_0^1 
    \likelihood(X) \ln \likelihood(X) \deriv X
    - \ln \evidence
\end{align}

Since $N$ is a discrete number that is typically
large, it is reasonable to assume that it follows
a Poisson distribution such that
\begin{equation}
    \mean{N} = \variance{N} \sim \frac{H}{\Delta \ln X}
\end{equation}
This leads to a rough uncertainty in $\ln \hat{\evidence}$ of
\begin{align}
    \stddev{\ln \hat{\evidence}}
    &\sim \stddev{\ln \hat{X}_N}
    \sim \stddev{\ln N} \, (\Delta \ln X) \nonumber \\
    &\sim \sqrt{H (\Delta \ln X)} = \sqrt{\frac{H}{K}}
    \label{eq:lnz_err_static}
\end{align}

We now extend this result to encompass a variable number of
live points $K_i$ at each iteration. We first rewrite
our estimate of the variance as
\begin{align}
    \variance{\ln \hat{\evidence}} 
    &= \variance{\sum_{i=1}^{N} \left( \ln \hat{\evidence}_i 
    - \ln \hat{\evidence}_{i-1} \right)}
    \equiv \variance{\sum_{i=1}^{N} \Delta \ln \hat{\evidence}_i}
    \nonumber \\
    &\approx \sum_{i=1}^{N} \variance{\Delta \ln \hat{\evidence}_i}
\end{align}
where the final approximation assumes the distribution of evidence updates
is independent at each iteration $i$ and $\ln \hat{\evidence}_0 = 0$.
If we further assume that the distribution of the
actual evidence estimates $\hat{Z}_i$ themselves are 
roughly independent at each iteration $i$ and that
the number of live points $K_i$ changes 
sufficiently slowly such that 
$\Delta \ln X_i \approx \Delta \ln X_{i-1}$,
we find
\begin{align}
    \variance{\Delta \ln \hat{\evidence}_i}
    &\approx \variance{\ln \hat{\evidence}_i}
    - \variance{\Delta \ln \hat{\evidence}_{i-1}} \nonumber \\
    &\sim (H_i - H_{i-1}) (\Delta \ln X_i)
    \equiv (\Delta H_i) (\Delta \ln X_i)
\end{align}
Substituting this in to our original expression
and taking $\Delta \ln X_i \approx 1/K_i$ then gives
a modified error estimate
\begin{equation}
    \stddev{\ln \hat{\evidence}} 
    \sim \sqrt{\sum_{i=1}^{N} \frac{\Delta H_i}{K_i}}
    \label{eq:lnz_err}
\end{equation}
While the modified estimator in equation \eqref{eq:lnz_err}
is less reliable that our original estimate, it
is somewhat reassuring that in the special case $K_1 = \dots = K_N = K$
it reduces to the original estimator derived in equation
\eqref{eq:lnz_err_static}.



\subsubsection{Sampling Uncertainties}
\label{subsubap:error_samp}

Unlike the statistical uncertainties on the prior volume estimator
$\hat{X}_i$, we do not have analytic expression or ways
to explicitly simulate from the distribution characterized by our sampling
uncertainties. In fact, it is doubtful we will 
ever have access to these except in special cases, since they rely
on having access to the distribution of
all possible paths (or varying lengths)
live points can take through the distribution over
the course of a Nested Sampling run.

This however, does not mean we cannot attempt to construct an
estimate of this distribution. To do this, we follow \citet{higson+17a}
and turn to bootstrapping, which serves as
a generic and robust tool for attempting
to simulate the impact of sampling uncertainties with limited
support \citep{efron79}. Since in most cases we have many thousands
of samples from our distribution and sample with $K>100$ live points,
Nested Sampling is almost always in a regime
where bootstrapping should be viable.

Naively, we might expect to simply be able simulate
values of, e.g., $\hat{\evidence}$ by just bootstrapping 
the underlying set of live points. However, this leads to three immediate
complications:
\begin{enumerate}
    \item This approach creates multiple samples at the same position.
    It is unclear how these points need to be ordered to assign them
    associated prior volumes.
    \item This approach conserves the total number of
    samples $N$, which clearly must be allowed to change if we really want
    to simulate from all possible live point paths (with varying path-lengths).
    \item This approach can leave out samples initially drawn from the
    prior. These points are crucial for establishing the normalization
    needed to estimate the evidence, and so removing them drastically
    distorts our evidence estimates.
\end{enumerate}
We address each of these in turn.

First, the ambiguous ordering, while at first glance a serious issue,
is actually a non-concern since the impact on any derived quantity
is actually completely insensitive to the ordering. For the evidence,
since the likelihood $\likelihood(\params)$ is identical among the
points, their contribution to $\hat{\evidence}$ will remain unchanged. 
Likewise, because they occupy the same position $\params$, 
their contribution to the posterior estimate $\hat{\posterior}(\params)$
is also unchanged. This implies that any ordering scheme (e.g., random)
will suffice.

To resolve the second issue, we now turn to the problem
of simulating all possible live point paths along with their possibly
varying path-lengths. Bootstrapping over all the samples by
construction destroys this information by ignoring the paths
of each individual live point. Analogous to the discussion in 
\S\ref{subap:nested_parallel}, we can characterize
these individual paths as being the collection of
$K$ lists of positions
$\params_1^{[j]} \rightarrow \dots \rightarrow \params_{N_j}^{[j]}$
traversed by each live point. Sampling from the space of
all possible live point paths thus is equivalent to bootstrapping
from these individual $K$ ``strands''
and then merging the $K$ resampled strands
$\lbrace \dots, \lbrace
\params_1^{[j']} \rightarrow \dots \rightarrow \params_{N_{j'}}^{[j']} 
\rbrace, \dots \rbrace $
into a new Nested Sampling run.

Unfortunately, this procedure still can run afoul of the third issue
when the number of live points $K_i$ is not constant. Going back to
the discussion in \S\ref{subap:nested_parallel} and
the Iterative Dynamic Nested Sampling scheme outlined in
Algorithm \ref{alg:dynamic_iter}, we see that increasing the
number of live points at some iteration $i > 1$ means that
those additional live points \textit{were sampled interior to the
prior} at some associated likelihood threshold $\likelihood(\params_i)$.
Since these live points provide no information about the overall normalization
(only the normalization relative to $\hat{X}_i$),
they are totally uninformative on their own when it comes to estimating
the evidence $\hat{\evidence}$. 

To account for this, we need to perform a
\textit{stratified} bootstrap over the set of $K_{\rm int}$
``interior'' strands (i.e. strands with starting positions
interior to the prior) and $K_{\rm anc}$ ``anchor'' strands
(i.e. strands sampled directly from the prior that ``anchor''
the interior strands). Once the set of $K_{\rm int}$ 
interior strands and $K_{\rm anc}$ strands have been resampled,
we can then merge the new collection into a new Nested Sampling
run. Following this scheme is then sufficient for
simulating the evidence $\hat{\tilde{\evidence}}$
and posterior $\hat{\tilde{\posterior}}(\params)$ estimates,
where we have used $\hat{\tilde{\evidence}}$ notation
to indicate a we used bootstrapping rather than 
prior volume simulation.

Note that one interesting corollary of our bootstrap estimates
is that we expect the total number of samples $\tilde{N}$ to change.
For a sufficient number of live points, this distribution is
likely to be roughly Poisson. Assuming that the associated
$\Delta \tilde{H}_i \approx \Delta H_i$ and $\tilde{K}_i \approx K_i$
from our bootstrapped Nested Sampling run are similar to the original,
this immediately leads us to an estimate of $\stddev{\ln\hat{\evidence}}$
identical to that in equation \eqref{eq:lnz_err}.
Although it has the exact same form, note that this error term
\textit{is completely independent} from the previous case.

\subsubsection{Combined Uncertainties}
\label{subsubap:error_comb}

The full uncertainties associated with a given Nested Sampling run
involve both the statistical uncertainties described in \S\ref{subsubap:error_stat}
and the sampling uncertainties described in \S\ref{subsubap:error_samp}.
Simulating from this combined error distribution is straightforward and
can be done by the following procedure:
\begin{enumerate}
    \item Resample the set of underlying $K_{\rm anc}$ anchor 
    and $K_{\rm int}$ interior strands using stratified bootstrap resampling.
    \item Merge the resampled strands into a single run.
    \item Simulate the values of the prior volumes.
\end{enumerate}
We can then calculate the evidence $\hat{\tilde{\evidence}}'$
and posterior $\hat{\tilde{\posterior}}'(\params)$ estimates accordingly.
The combined uncertainty on the evidence that we estimate from
both sources is then roughly
\begin{equation}
    \stddev{\ln \hat{\evidence}} 
    \sim \sqrt{2 \sum_{i=1}^{N} \frac{\Delta H_i}{K_i}}
    \label{eq:lnz_err_tot}
\end{equation}
based on the identical error estimates derived in
\S\ref{subsubap:error_stat} and \S\ref{subsubap:error_samp}.


\bsp	
\label{lastpage}
\end{document}